%% file: main.tex
\documentclass[acmtog, nonacm]{acmart}

\AtBeginDocument{%
  }

\setcitestyle{nosort}

\usepackage{content/macros}
\usepackage{siunitx}
\usepackage[table]{xcolor}

\setcopyright{none}
\renewcommand\footnotetextcopyrightpermission[1]{}
\authorsaddresses{}

\begin{document}

\title[Monte Carlo Steklov Operators]{Monte Carlo Steklov Operators \\
for Large-Scale Geometry Processing in the Wild}

\author{Arman Maesumi}
\authornote{These authors contributed equally to this work.}
\email{arman.maesumi@gmail.com}
\orcid{0000-0001-7898-8061}
\affiliation{%
  \institution{Brown University}
  \city{Providence}
  \country{USA}}

\author{Tanish Makadia}
\authornotemark[1]
\email{tanish_makadia@brown.edu}
\affiliation{%
  \institution{Brown University}
  \city{Providence}
  \country{USA}}

\author{Aruna Anderson}
\authornote{These authors contributed equally to this work as co-second authors.}
\email{asunander@hotmail.com}
\affiliation{%
  \institution{Loyola Marymount University}
  \city{Los Angeles}
  \country{USA}}

\author{Oras Phongpanangam}
\authornotemark[2]
\email{panangam@gmail.com}
\affiliation{%
  \institution{Brown University}
  \city{Providence}
  \country{USA}}

\author{Justin Solomon}
\email{jsolomon@mit.edu}
\affiliation{%
  \institution{Massachusetts Institute of Technology}
  \city{Boston}
  \country{USA}}

\author{Daniel Ritchie}
\email{daniel_ritchie@brown.edu}
\affiliation{%
  \institution{Brown University}
  \city{Providence}
  \country{USA}}

\renewcommand{\shortauthors}{Maesumi and Makadia et al.}

\begin{abstract}
  \input{content/abstract}
\end{abstract}




% Teaser figure
\begin{teaserfigure}
  \includegraphics[width=\textwidth]{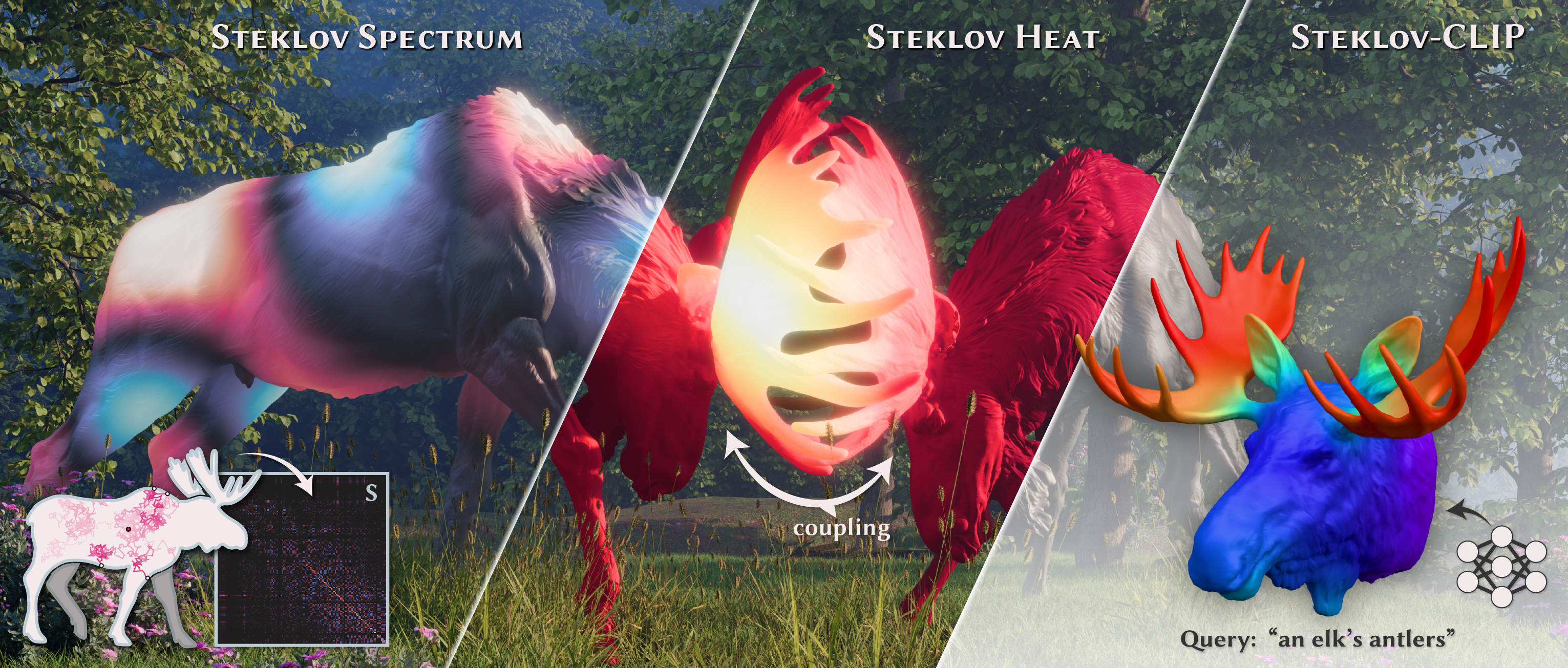}
  \caption{We present a Monte Carlo method for estimating the Dirichlet-to-Neumann (DtN) operator and its associated Steklov eigenmodes, enabling fast and robust volumetric analysis of \emph{in-the-wild} shapes. \textit{Left:} We formulate a Brownian process that resolves entries of a discrete DtN operator---the 87th eigenvector of this operator is visualized on the moose, which contains 2 million faces. \textit{Center:} Our method applies to both \emph{interior} and \emph{exterior} operators---the latter encodes coupling behavior between disconnected geometries. \textit{Right:} Using our estimated Steklov operators, we train a mesh-based CLIP model that learns meaningful global and dense shape representations, which is able to isolate semantic components via text, e.g. \texttt{``an elk's antlers.''}}
  \Description{Teaser description for accessibility.}
\end{teaserfigure}

\maketitle

%% ----- Body -----
\input{content/sections/intro}
\input{content/sections/related-work}
\input{content/sections/preliminaries}
\input{content/sections/method}
\input{content/sections/implementation}
\input{content/sections/evaluation}

\input{content/sections/applications}
\input{content/sections/conclusion}

%%
%% If your work needs an appendix, add it before the
%% "\end{document}" command at the conclusion of your source document.
%%
%% Start the appendix with the "appendix" command:

\begin{acks}
The authors are deeply grateful to Noam Aigerman for his guidance and involvement in foundational work from which the initial ideas in this paper grew. We also thank Nicole Ge for her involvement in early experimentation. Finally, the authors are thankful for fruitful discussions with Nick Sharp, Sina Nabizadeh, Ana Dodik, Derek Liu, Keenan Crane, and Siddhartha Chaudhuri.
\end{acks}

\bibliographystyle{ACM-Reference-Format}
\bibliography{content/references}

\appendix
\input{content/sections/supplemental}

\end{document}

%% file: content/abstract.tex
\label{abstract}
Intrinsic methods fill the default toolbox for geometry processing on meshes. Intrinsic operators, in particular the Laplacian, underlie methods that require invariance to isometry and have hence been employed in many algorithms for shape analysis, learning, and editing.
However, intrinsic methods are predicated on assumptions that quickly become brittle when working with in-the-wild geometry, where (i) mesh quality is not guaranteed, and (ii) many meshes are modeled with multiple connected components.
In such settings, volumetric constructions are better-defined, since restrictions on surface topology can be relaxed.
This paper presents a Monte Carlo method for estimating the Dirichlet-to-Neumann (DtN) operator---a boundary-to-boundary volumetric operator---and its associated Steklov eigenmodes. We build on recent developments in Monte Carlo geometry processing by casting this boundary operator itself as the subject of estimation. 
The DtN operator, defined through a volumetric stochastic process, is then generalized to the exterior domain, where it couples disconnected components through the surrounding ambient space. 
We show that our method is orders of magnitude faster than existing boundary-element approaches for computing Steklov spectra while remaining robust to poor triangulations, high-resolution meshes, and multi-component geometry. To demonstrate this scalability, we compute interior and exterior Steklov eigenspectra for approximately 450,000 shapes from the uncurated Objaverse dataset. We incorporate these operators into Steklov-CLIP, a mesh-based neural network that uses volumetric spectral operators for large-scale contrastive 3D representation learning. The resulting network learns semantically meaningful global and dense shape representations, illustrating that geometrically-principled volumetric operators can be made practical at the scale of modern 3D datasets. Code: \href{https://github.com/mc-steklov/mc-steklov}{\color{blue} https://github.com/mc-steklov/mc-steklov}

%% file: content/sections/intro.tex
\section{Introduction}
\label{sec:intro}

The intrinsic Laplacian operator has long served as the workhorse of geometry processing. Because it depends only on the metric of a surface, it is invariant to isometric deformation, admits sparse local discretizations, and underpins methods for shape editing~\cite{sorkine2007rigid, jacobson2011bounded}, parametrization~\cite{desbrun2002intrinsic, sawhney2017boundaryFlat, levy2023least}, correspondence~\cite{ovsjanikov2012functional, litany2017deep}, decomposition~\cite{huang2009shape, reuter2010hierarchical}, and geometric deep learning~\cite{smirnov2021hodgenet,sharp2022diffusionnet,wiersma2022deltaconv,maesumi2025poissonnet}.

The assumptions behind many intrinsic pipelines, however, have become restrictive in the age of contemporary geometry processing, which typically features large, minimally-curated shape collections whose meshes vary substantially in quality~\cite{deitke2023objaverse}.
The discretized operators (e.g., the cotangent Laplacian matrix) employed by intrinsic methods often expect clean, single-component, manifold meshes with reasonable element quality, and these operators are highly sensitive to spurious changes in topology that are incidental to how the mesh was modeled.

For such data, an extrinsic or volumetric viewpoint can be more natural. The occupied volume of a shape is frequently a more stable geometric signal than the surface induced by scanning, reconstruction, 3D generative models, or manual authoring; moreover, volumetric operators capture structure that intrinsic methods necessarily ignore. Indeed, volumetric and extrinsic boundary-based methods have proven useful in a range of shape-analysis settings (see discussion in Section \ref{sec:related-work}).%dirac operator, laplacian with mean curvature terms, dtn, shell-based shape difference paper, ...

Among extrinsic operators, the \emph{Dirichlet-to-Neumann (DtN) operator} is particularly appealing for its applications to geometry processing \cite{wang-2017-steklov}.  Given a scalar function on a closed surface, the DtN operator returns the normal flux of its (volumetric) harmonic extension through the boundary. Although it acts purely as a boundary-to-boundary operator, it encodes volumetric geometry: its eigenpairs are the \emph{Steklov modes}, which provide an extrinsic spectral description of shape. \citet{wang-2017-steklov} demonstrated the practical utility of this operator for geometry processing, showing that Steklov spectra stably introduce extrinsic information to geometry processing pipelines while avoiding tetrahedralization via a boundary element (BEM) formulation.

\begin{figure}
    \centering
    \includegraphics[width=\linewidth , trim=0 0.3cm 0 0]{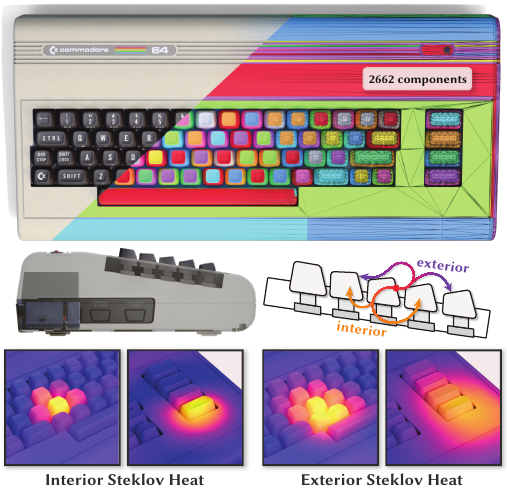}
    \caption{The heat equations governed by interior and exterior Steklov eigenspectra exhibit non-trivial dynamics. We evolve a Dirac source distribution localized to a single keycap, and see that the interior heat evolves through the hollowed area beneath each cap, leaking into adjacent ones. The exterior heat evolves with a wider profile through the ambient space surrounding the keys. We note that the Commodore 64 keyboard has 2662 connected components that are triangulated very poorly, yet our method's estimated Steklov eigenmodes nevertheless give high-quality results.}
    \label{fig:multi_component}
\end{figure}

Scalability and robustness are key obstacles for volumetric methods in general. Classical finite-element and boundary-element methods for volumetric operators are substantially more expensive than the sparse intrinsic pipelines that dominate surface processing. Even when tetrahedralization is avoided, discretizations of volumetric operators are typically dense, and hence computing Steklov eigenspectra can require on the order of tens of minutes on meshes that are modest by the standards of current graphics datasets. Methods that rely on tetrahedralization similarly are accompanied by exorbitant up-front preprocessing costs that may fail or not fully preserve the original surface mesh (see Figure 4 and Table 1 from \citet{dodik2025biharmonic}). Even when one accepts these drawbacks, these pipelines may not terminate successfully, e.g.\ due to sensitivity to poor element quality or quadrature. These bottlenecks preclude large-scale volumetric analysis, especially for datasets containing high-resolution meshes with poor element quality.

Meanwhile, Monte Carlo methods have emerged as an attractive approach for solving volumetric elliptic boundary value problems in graphics~\cite{sawhney2020mc,sawhney2023stars}. These works extend the Walk-on-Spheres (WoS) method~\cite{muller-1956-wos}, which estimates the solution of Laplace's equation by simulating Brownian motion and does not require a volumetric mesh or dense intermediate constructions. WoS is progressive, naturally parallelizable, and comparatively insensitive to poor element quality or local geometric degeneracies~\cite{sawhney2020mc}. Most of these Monte Carlo methods, however, estimate \emph{solutions} to PDEs at selected query points. For shape analysis, the object of interest is different: we would like to estimate the \emph{boundary operator} itself, enabling downstream geometry processing algorithms.

In this paper, we introduce a Monte Carlo method for estimating Dirichlet-to-Neumann operators, and in particular their associated Steklov eigenspectra. We first demonstrate that a na\"ive Monte Carlo estimator for these operators results in shortcomings that mirror those of previous volumetric methods: dense linear algebra and compromised structure due to poor sampling dynamics (e.g., high variance).
As an alternative, we use the Beurling-Deny Formula~\cite{beurling-deny-1958-ice} to rewrite DtN operators, uncovering a Monte Carlo estimator that produces a positive semidefinite matrix from any finite number of samples. We decompose the rewritten integral forms of these operators into analytic and estimated parts, reducing sampling variance. We further observe that Steklov eigenspectra  are well-approximated in a compact functional basis on the surface, bypassing dense intermediate constructions. Our estimators are implemented in CUDA and yield a highly-practical pipeline for using DtN operators even on meshes with millions of elements or poor surface discretizations.

Further, by applying the DtN operator (and our estimator) to \emph{exterior} domains, we obtain an operator with an appealing property for geometry processing: unlike local surface-based operators, which act independently on disconnected components, the exterior DtN operator \emph{couples} boundary regions that are topologically disconnected on the mesh but interact through the ambient space (see Figures \ref{fig:multi_component} and \ref{fig:torus}). We demonstrate the utility of this property on shapes with disconnected components and cavities, where it enables information to be propagated across disjoint regions.

\smallskip
\noindent 
We summarize our contributions as follows:
\begin{itemize}
    \item We derive Monte Carlo estimators for the interior and exterior Dirichlet-to-Neumann operators on triangle meshes and estimate their Steklov eigenspectra efficiently without requiring a tetrahedral mesh or large dense intermediate matrices.
    \item We demonstrate the scalability and robustness of our approach by computing interior and exterior Steklov eigenspectra across ${\sim}450$K shapes from Objaverse, far exceeding what is possible with existing spectral volumetric methods. 
    \item We incorporate these operators into a neural network and perform mesh-based 3D representation learning via contrastive training. We demonstrate that our network, Steklov-CLIP, learns rich representations on meshes that are aligned with an existing text-image contrastive model (i.e. CLIP). The network is further finetuned to produce dense representations that facilitate spatial queries (e.g. zero-shot semantic selection of parts).
\end{itemize}

%% file: content/sections/related-work.tex
\begin{figure}
    \centering
    \includegraphics[width=\linewidth , trim=0 0.3cm 0 0]{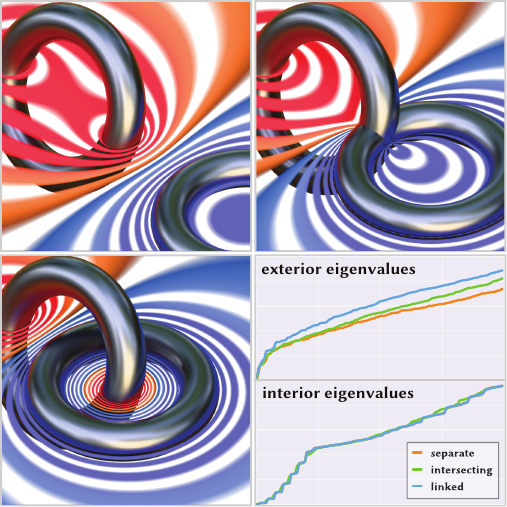}
    \caption{The exterior DtN operator---being defined through the exterior harmonic extension (visualized as iso-potentials around each torus)---is sensitive to the relative positions and orientations of objects in a scene. By contrast, the interior DtN operator is relatively stable to such changes, due to the shapes being closed. This difference in behavior can be seen in their associated eigenspectra (bottom right).}
    \label{fig:torus}
\end{figure}

\section{Related Work}
\label{sec:related-work}

\paragraph*{Extrinsic Geometry Processing.}
Geometry processing algorithms commonly build on the Laplace-Beltrami operator and its eigenspectrum due to its invariance to isometries. This invariance is also a limitation: a purely intrinsic operator cannot distinguish shapes that share the same metric---e.g. a human mesh in different poses. A broad line of work therefore replaces or augments the Laplacian with operators that expose extrinsic information; see the comprehensive survey of \citet{wang2019survey} for further background. Representative examples include discrete Dirac operators \cite{liu2017dirac,ye2018intrinsicExtrinsicDirac} and functional shape-difference methods that separate intrinsic and extrinsic quantities \cite{corman2017functional}. Closest to our setting is Steklov geometry processing \cite{wang-2017-steklov}, which uses the Dirichlet-to-Neumann operator---in particular its eigenspectrum---as an extrinsic spectral operator. \citet{wang-2017-steklov} demonstrated that the Steklov eigenmodes can be inserted into standard spectral geometry pipelines, but they rely on a boundary element method that limits scalability to large meshes and is brittle in the face of poorly triangulated ones. Our work builds on this boundary-operator viewpoint, but instead estimates interior and exterior DtN operators by stochastic (Monte Carlo) sampling, which allows us to apply this machinery to large meshes with degenerate triangulations, and is orders of magnitude faster than existing volumetric spectral methods.

\paragraph*{Spectral Geometry.}
Spectral shape analysis represents geometry through eigenvalues and eigenfunctions of geometric operators. Seminal applications include spectral shape descriptors built from the Laplace-Beltrami eigenmodes \cite{sun2009hks,bronstein2010scaleInvHKS,aubry2011WKS}; functional maps, which represent dense correspondence as compact linear operators in reduced eigenbases \cite{ovsjanikov2012functional}, as well as later learning-based functional map pipelines \cite{litany2017deep}; finally, select methods for geometric learning define feature transformations using filters defined in the spectral domain \cite{smirnov2021hodgenet,sharp2022diffusionnet, gao2024intrinsic}. The Steklov eigenspectrum is attractive because its eigenfunctions live on the surface, while the underlying operator is induced by a \emph{volumetric} harmonic extension. The exterior variant of this operator further changes the information encoded by its associated spectrum since the relevant harmonic process takes place in the ambient complement of the volume. Through this process, the exterior Steklov spectrum couples disconnected geometry based on harmonic accessibility through the surrounding space. Components that are topologically disjoint may nevertheless interact strongly when Brownian motions through the exterior domain readily have access to them, while intervening ``barriers'' can attenuate this interaction, making the process---and hence the spectrum---highly informative of the geometry at hand.

\paragraph*{Monte Carlo PDE}
The stochastic foundations of our method go back to Kakutani's representation of harmonic functions through the lens of Brownian motion and Muller's Walk-on-Spheres algorithm for the Dirichlet problem \cite{kakutani-1944-brownian,muller-1956-wos}. Recent work in graphics has revived these ideas to motivate grid-free solvers for elliptic PDEs on complex geometry. In particular, \citet{sawhney2020mc} introduced Muller's Walk-on-Spheres method to the graphics community and extended it to include efficient estimation of solution derivatives and variance reduction techniques. Subsequent work expanded on this methodology to PDEs with spatially-varying coefficients \cite{sawhney2022svcoeff}, infinite domains via the Kelvin transform \cite{nabizadeh-2021-kelvin}, mixed Neumann/Dirichlet and Robin boundary conditions \cite{sawhney2023stars,miller2024walkin}, and further variance reduction techniques have been considered \cite{miller2023BVC, neuralControlVariates,huang2025guiding}, along with methods for fast recomputation using local solution operators~\cite{wosSubDomain}. Boundary-integral Monte Carlo methods provide another perspective on this stochastic framework by considering random walks directly on surfaces \cite{karlovivc1994walkonboundary, sugimoto2023walkonboundary}. Differential Monte Carlo solvers estimate sensitivities or normal derivatives of PDE solutions, including derivatives with respect to boundary data or shape \cite{miller2024differential,yu-2024-tangent-ball}. The estimators introduced in this paper address a complementary goal---rather than estimating the solution, gradient, or derivative of a PDE at selected query points, we instead estimate \emph{boundary operators} themselves and extract eigenspectra from them, to enable fast and robust volumetric spectral geometry processing.

%% file: content/sections/preliminaries.tex
\begin{figure}
    \centering
    \includegraphics[width=\linewidth , trim=0 0.3cm 0 0]{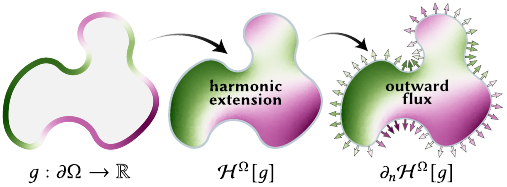}
    \caption{The Dirichlet-to-Neumann operator maps scalar boundary functions, $g$, to the outward normal derivative of their harmonic extension.}
    \label{fig:dtn_illustrated}
\end{figure}

\section{Preliminaries}
\label{sec:preliminaries}

Our goal is to devise a Monte Carlo estimator for the Dirichlet-to-Neumann operator that enables rapid and robust approximation of its eigenspectrum. We begin with elementary definitions from PDEs that are necessary for defining this operator. We briefly introduce stochastic perspectives, which motivate Monte Carlo methods for such problems. Finally, we introduce the Dirichlet-to-Neumann operator, along with its corresponding Steklov eigenproblem.

\paragraph*{Dirichlet Problem.} Let $\vol\subset\R^3$ be a bounded volumetric domain with smooth boundary $\boundary$. Given a scalar-valued function $g: \boundary\to\R$, the \emph{Dirichlet Problem} seeks a harmonic function $u:\vol\to\R$ on the interior that equals $g$ on the surface:
\begin{align}
    \label{eq:dirichlet-problem}
    \left\{
    \begin{aligned}
        \Delta u &= 0 & &\text{in } \vol,\\
        u &= g & &\text{on } \boundary,
    \end{aligned}
    \right.
\end{align}
where $\Delta$ is the Laplacian operator. The solution $u$ uniquely extends $g$ to the volume, and hence it is referred to as the \emph{harmonic extension} of $g$. We denote the harmonic extension operator as $\H^\vol[g]=u$.

\paragraph*{Harmonic Measure and the Poisson Kernel.}

Let $x\in\vol$ be a point in the domain's interior. The harmonic extension operator $\H^\vol[g]$ defines a probability measure $\omega^{\vol}_x$ over the boundary by
\begin{align}
    \label{eq:harmonic-measure}
    u(x) &= \int_{\boundary} g(s)\ d\omega^\vol_x(s).
\end{align}
Here, $\omega_x^\vol(s)$ is the \emph{harmonic measure} of a surface point $s\in\boundary$ with respect to the interior point $x$. Since harmonic measure and surface area measure $\sigma$ are both defined over the same measurable space $\boundary$, the Radon-Nikodym Theorem gives a density $\P_x^\vol = d\omega_x^\vol / d\sigma$, called the \emph{Poisson kernel}. Thus, for any surface patch $A\subseteq\boundary$,
\begin{align}
    \label{eq:radon-nikodym-derivative}
    \omega_x^\vol(A) &= \int_{s\in A} \P_x^\vol(s)\ d\sigma(s).
\end{align}
Hence, $\P_x^\vol$ is the \emph{Radon-Nikodym derivative} of harmonic measure with respect to surface area. Combining this definition with \refeq{eq:harmonic-measure}, the Poisson kernel's utility comes from the fact that it solves the Dirichlet problem with a boundary integral
\begin{align}
    \label{eq:poisson-kernel-analytic}
    u(x) &= \int_{\boundary} g(s)\,\P^\vol_x(s)\ d\sigma(s).
\end{align}

A probabilistic interpretation of the Poisson kernel was given by \citet{kakutani-1944-brownian}, who showed that if $B_t$ is a Brownian motion started at $x$ and $\tau=\inf\{t\mid B_t\notin\vol\}$ is the first time $B_t$ exits the volume, then $\H^\vol[g]$ is given by
\begin{align}
    \label{eq:poisson-kernel-probabilistic}
    u(x) &= \Exp[g(B_\tau)],
\end{align}
proving that $\omega^\vol_x$ is a probability distribution over exit points $B_\tau$ with density $\P^\vol_x$. To better represent this perspective, we denote the Poisson kernel as $\P^\vol(x\to s)$, giving the probability density that a Brownian motion originating from $x\in\vol$ first hits the boundary at $s\in\boundary$.

\paragraph*{Connection to Green's Function.} The Poisson kernel has an equivalent characterization through the \emph{Green's function} of the Laplacian operator. The Green's function $G^\vol(x,s)$ is defined by
\begin{align}
    \label{eq:greens-func}
    \Delta_x G^\vol(x,s) &= \delta(x-s).
\end{align}
It describes the system's response at $x\in\vol$ to a point-source placed at $s\in\vol$. The Poisson kernel is recovered by taking the normal derivative of $G^\vol$ at the boundary,
\begin{align}
    \label{eq:poisson-kernel-to-greens-func}
    \P^\vol(x\to s) &= \partial_{n_s}G^\vol(x,s),
\end{align}
where $n_s$ is the outward unit normal at $s\in\boundary$.

\paragraph*{Dirichlet-to-Neumann Operator.} The Poisson integral (\refeq{eq:poisson-kernel-analytic}) maps boundary values $g$ to their harmonic extension $u$. Similarly, the \emph{Dirichlet-to-Neumann} (DtN) \emph{operator} 
\begin{align}
    \label{eq:dtn}
    \dtn: H^\frac{1}{2}(\boundary)\to H^{-\frac{1}{2}}(\boundary)\quad\text{where}\quad
    g\mapsto\partial_n\H^\vol[g]
\end{align}
maps $g$ to the normal flux of $u$. Here, $H^k$ denotes the Sobolev space of order $k$. This boundary-to-boundary operator can be viewed as a composition; first take the harmonic extension of Dirichlet data $g$ to obtain $u$, and then take the outward normal derivative $\partial_nu$ to obtain Neumann data (as illustrated in Figure \ref{fig:dtn_illustrated}). Because $\dtn$ maps between different classes of boundary conditions, it falls into a broader family of operators known as \emph{Poincaré-Steklov operators} \cite{agoshkov1988poincare}.

\paragraph*{Steklov Eigenspectrum.} The \emph{Steklov eigenproblem} seeks nonzero
boundary data $\psi$ whose harmonic extension has normal derivative proportional
to its boundary trace. In other words, if $u=\H^\Omega[\psi]$ satisfies
\begin{align}
    \label{eq:steklov-eigenproblem}
    \left\{
    \begin{aligned}
        \Delta u &= 0 & &\text{in } \vol,\\
        \partial_n u &= \lambda u & &\text{on }\boundary,
    \end{aligned}
    \right.
\end{align}
then $\dtn \psi= \lambda \psi$. Thus $\psi$ is an eigenfunction of $\dtn$ with
eigenvalue $\lambda$. We refer to the corresponding eigenpairs $(\lambda,\psi)$
as the \emph{Steklov eigenmodes} of $\dtn$. The primary interest of this paper lies in approximating this spectrum efficiently to enable downstream applications. We further discuss the utility of the Steklov eigenspectrum in Section \ref{sec:related-work}.

\paragraph*{Properties of DtN}
The DtN operator has a natural bilinear form that reveals its core structure. For two boundary functions $f,g\in H^\frac{1}{2}(\boundary)$, define
\begin{align}
    \label{eq:dirichlet-form-dtn-boundary}
    \E[f, g] &:= \int_{\boundary} f(s)(\dtn g)(s)\ ds.
\end{align}
Applying Green's first identity to the harmonic extensions $u_f=\H^\vol[f]$ and $u_g=\H^\vol[g]$ transforms this boundary integral into a volume integral,
\begin{align}
    \label{eq:dirichlet-form-dtn-intermediate}
    \E[f,g] &= \int_{\boundary} f(\partial_n u_g)\ ds \\
    \label{eq:dirichlet-form-dtn-volume}
    &= \int_\vol \nabla u_f\cdot \nabla u_g\ dV.
\end{align}
The right-hand side is the \emph{Dirichlet energy} inner product of two harmonic extensions. This identity has two immediate consequences.
\begin{enumerate}
    \item \textbf{Symmetry.} Since $\E[f,g]=\E[g,f]$, we know that $\dtn$ is a self-adjoint operator.

    \smallskip
    \item \textbf{PSD.} Setting $f=g$ gives $\E[f,f]=\int_\vol |\nabla u_f|^2\ dV\geq 0$. Hence, $\dtn$ is positive semidefinite.
\end{enumerate}
The latter property in particular is central to the development of our core method in Section \ref{sec:monte-carlo}.

\paragraph*{Walk-on-Spheres.}
The probabilistic identity in \refeq{eq:poisson-kernel-probabilistic} suggests a method for estimating the harmonic extension of boundary data. By simulating a Brownian motion from $x$ and recording where it first hits $\boundary$, we can sample exit points $B_\tau\sim\omega_x^\vol$ from a distribution matching the harmonic measure. Directly tracing a continuous Brownian path, however, is intractable in practice. 

As an efficient alternative, Walk-on-Spheres (WoS) \cite{muller-1956-wos} leverages the \emph{Mean Value Property} of harmonic functions. For any ball $B(x)$ centered at $x$, a Brownian motion started at $x$ has the same probability of exiting at any point on $\partial B(x)$. WoS exploits this fact by repeatedly constructing the largest ball that is fully contained in $\vol$ and jumping to a uniformly random point on its surface. The walk terminates once it arrives within a small tolerance $\epsilon>0$ of $\boundary$, after which the nearest boundary point is returned as the exit location. WoS converges to $\boundary$ in $O(\log 1/\epsilon)$ time for sufficiently smooth boundaries \cite{binder-braverman-2012-wos}.

%% file: content/sections/method.tex
\section{Monte Carlo Estimation of DtN Operators}
\label{sec:monte-carlo}

Our goal is to recover the low-frequency Steklov eigenspectrum of a triangle mesh $\mathcal{M}$ with vertices $\*V$ and faces $\*F$, which approximates an underlying boundary $\boundary$. Rather than obtaining the spectrum directly, we estimate---with Monte Carlo---the Dirichlet-to-Neumann operator, whose leading eigenpairs encode the Steklov modes of interest. We begin by deriving a straightforward Monte Carlo estimator w.r.t.\ a discrete boundary mesh to make the basic idea concrete. This na\"ive estimator is impractical, so we propose an alternative formulation in terms of a jump-kernel identity that yields a practical DtN estimator, and hence the Steklov eigenspectrum. Our estimator yields a discrete operator that is guaranteed to be PSD (regardless of Monte Carlo noise) and avoids materializing large, dense intermediate matrices. We additionally extend to the exterior DtN problem, which is accelerated with a Kelvin transform.
\begin{prettyalgorithm}[t]{alg:naive-estimator}{Na\"ive Estimator for Interior DtN}
\begin{algorithmic}
  \Require Boundary mesh $\mathcal{M}$, inset distance $\inset$, sample count $N$
  \Ensure Monte Carlo estimate $\widehat{\*S}\in\R^{V\times V}$ of the DtN stiffness matrix
  \State $\widehat{\*S} \gets \*0 \in \R^{V\times V}$
  \For{$m = 1,\ldots,N$}
    \State $s,\ \mathrm{tri}_s \gets$ \Call{UniformSampleMesh}{$\mathcal{M}$}
    \State $x \gets s - \inset\, n_s$\Comment{Inset evaluation point}
    \State $t,\ \mathrm{tri}_t \gets$ \Call{WalkOnSpheres}{$x$}\Comment{Sampled termination point}
    \State $b_s \gets$ \Call{BarycentricCoords}{$s,\mathrm{tri}_s$}\Comment{Known, \texttt{(V,)}}
    \State $b_t \gets$ \Call{BarycentricCoords}{$t,\mathrm{tri}_t$}\Comment{Sampled, \texttt{(V,)}}
    \State $\widehat{\*S} \gets \widehat{\*S}
      + \dfrac{|\boundary|}{N\inset}\,
      b_s \bigl(b_s - b_t\bigr)^\T$\Comment{Sparse rank-1 update, \texttt{(V,V)}}
  \EndFor
  \State $\widehat{\*S} \gets \frac{1}{2}(\widehat{\*S} + \widehat{\*S}^{\,\T})$\Comment{Make empirical operator symmetric}
  \State \Return $\widehat{\*S}$
\end{algorithmic}
\end{prettyalgorithm}
\subsection{A Na\"ive Estimator}
\label{par:naive-estimator}
The most straightforward approach to estimate $\Lambda$ with Monte Carlo is to write the directional derivative, $\partial_n$, in \refeq{eq:dirichlet-form-dtn-intermediate} as a boundary limit
\begin{align}
    \label{eq:finite-difference-boundary-limit}
    \E[f,g] &= \int_{\boundary} f(s) \cdot \left(\lim_{\inset\to 0} \frac{g(s) - u_g(s-\inset n_s)}{\inset}\right)\ ds.
\end{align}
We write the Monte Carlo estimator in terms of a boundary point $s$ sampled from the uniform distribution $U(\boundary)$. From $s$, we estimate the normal derivative of the harmonic extension by launching a WoS sample from an inset point $x(s) = s-\inset n_s$, where $n_s$ is the outward surface normal and $\inset$ is a chosen distance used to approximate the derivative. Then, the bilinear form can be written as the expectation
\begin{equation}
    \label{eq:finite-difference-estimator}
    \E[f,g] \approx \Exp_{s\sim U(\boundary),t\sim \omega^\Omega_{x(s)}}\left[ |\boundary|\: f(s)\: \frac{g(s) - g(t)}{\inset}\right].
\end{equation}

To discretize this expression, let $\basis=\{\phi_1,\phi_2,\ldots,\phi_V\}$ denote the standard piecewise-linear hat basis on $\mathcal{M}$. The discrete bilinear form is represented by the stiffness matrix $\*S\in\R^{V\times V}$ with entries
\begin{equation}
    \label{eq:naive-stiffness-entry}
    S_{ij}
    \approx
    \frac{|\boundary|}{\inset}\;
    \Exp_{s\sim U(\boundary),\,t\sim \omega^\Omega_{x(s)}}
    \!\left[
        \phi_i(s)\bigl(\phi_j(s)-\phi_j(t)\bigr)
    \right].
\end{equation}
If $b(p)=(\phi_1(p),\ldots,\phi_V(p))^\T$ denotes the vector of hat-function values at a boundary point $p$, then we can equivalently write the following expression for the entire stiffness matrix $\*S$ 
\begin{equation}
    \label{eq:naive-stiffness-matrix}
    \*S
    \approx
    \frac{|\boundary|}{\inset}\;
    \Exp_{s\sim U(\boundary),\,t\sim \omega^\Omega_{x(s)}}
    \!\left[
        b(s)\bigl(b(s)-b(t)\bigr)^\T
    \right].
\end{equation}
Here $b(p)$ is the sparse vector of barycentric coordinates of $p$ in its corresponding triangle. Thus, a WoS sample contributes a sparse rank-one update supported only on the vertices of the source and exit triangles. Averaging updates over $N$ samples yields the empirical operator $\widehat{\*S}$. We summarize this procedure in Algorithm \ref{alg:naive-estimator}.

While conceptually straightforward, this estimator suffers from two deficiencies that render it impractical for downstream tasks:

\medskip
\begin{enumerate}
    \item \textbf{PSD only in expectation.} \label{issue:psd} 
    The Steklov matrix $\stiffness$ is symmetric positive semidefinite (see \refeq{eq:dirichlet-form-dtn-volume}); however, the straightforward estimator in \refeq{eq:naive-stiffness-matrix} is only PSD \emph{in expectation}, specifically because the outer product is taken over generally distinct vectors. A finite-sample estimate can easily violate this condition, and in practice the estimated matrix frequently has negative eigenvalues, which corrupts the behavior of downstream algorithms.
    
    \medskip
    \item \textbf{High variance.} \label{issue:variance} 
    This estimator inherits a well-known weakness of Monte Carlo estimators for differential quantities: approximating a normal derivative by a finite-difference quotient is noisy, and the variance blows up as $\inset$ is decreased to reduce bias. In addition, choosing a globally-valid inset distance $\inset$ is itself delicate on meshes with thin features, where even a modest inset can leave the domain and lead to incorrect MC samples.
\end{enumerate}

\medskip\noindent
The first issue is a fundamental shortcoming of the bilinear form (\refeq{eq:finite-difference-boundary-limit}) from which the na\"ive estimator was derived. Below, we proceed with a key insight of this paper: the estimator can instead be derived from an alternative representation of the DtN operator, an expectation over rank-1 PSD outer products, yielding a PSD estimate. Further, we show that this representation can be decomposed into analytical and empirical parts, substantially reducing variance.

\begin{figure}
    \centering
    \includegraphics[width=\linewidth , trim=0 0.3cm 0 0]{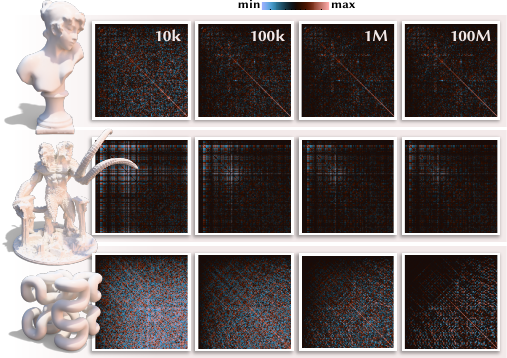}
    \caption{Progression of our interior DtN estimator over several Monte Carlo sample counts. Operators are represented in a Galerkin basis of size 128.}
    \label{fig:operator_convergence}
\end{figure}

\subsection{A Jump-Kernel Decomposition}
\label{sec:improved-beurling-deny-estimator}
The Beurling-Deny Theorem \cite{beurling-deny-1958-ice} asserts that regular\footnote{\emph{Regular} implies additional constraints on the Dirichlet form's domain and underlying topology, see Section 1.1 of \citet{fukushima-1994-dtn}.} Dirichlet forms decompose into three canonical parts: a \emph{strongly-local} (diffusion) term, a \emph{jump} term, and a \emph{killing} term. For the DtN Dirichlet form $\E$, which arises as the trace form of volumetric Dirichlet energy (\refeq{eq:dirichlet-form-dtn-boundary}), only the jump component survives. Concretely, as shown by \citet[Eq. 5.8.4]{chen-fukushima-2011-dtn},
there exists a nonnegative symmetric kernel $\J^\vol: \boundary\times\boundary\to\R_{\geq0}$ such that
\begin{align}
    \label{eq:beurling-deny-bilinear}
    \E[f, g] &= \frac{1}{2}\iint\limits_{\boundary\times\boundary} \big(f(s)-f(t)\big)\big(g(s)-g(t)\big)\,\J^\vol(s,t)\ ds\ dt.
\end{align}
This identity
decomposes the DtN's Dirichlet energy form---originally a volume integral (\refeq{eq:dirichlet-form-dtn-volume})---into a symmetric boundary-to-boundary process that we can sample. We call $\J^\vol$ the \emph{jump kernel} of $\dtn$, since it governs how the associated boundary process jumps between distant surface points. Translating the work of \citet[Eq. 5.8.2]{chen-fukushima-2011-dtn} into our conventions, the jump kernel,
\begin{align}
    \label{eq:jump-kernel-poisson}
    \J^\vol(s,t) &= -\partial_{n_s}\P^\vol(s\to t),
\end{align}
is the inward normal derivative of the Poisson kernel. This equation should be understood as a boundary limit, where 
\begin{align}
    -\partial_{n_s}P^\vol(s\to t):=\lim_{\epsilon\to 0}\frac{1}{\epsilon}P^\vol (s-\epsilon n_s\to t)
\end{align} 
because $G^\vol(s,t)$ and $\P^\vol(s\to t)$ are both zero when $s,t\in\boundary$, and $\P^\vol(s\to t)$ has a singularity when $s=t$.

\begin{figure}
    \centering
    \includegraphics[width=\linewidth , trim=0 0.3cm 0 0]{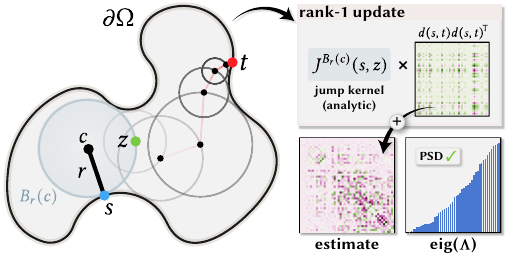}
    \caption{\textbf{Our final interior Dirichlet-to-Neumann estimator.} A point $s\sim U(\boundary)$ is sampled uniformly from the surface, from which we compute its largest tangent ball $B_r(c)$ with radius $r$ centered at $c$. The integral decomposition in Section \ref{sec:improved-beurling-deny-estimator} allows us to write the DtN estimator w.r.t. points $z$ drawn uniformly on $\partial B_r(c)$, effectively factoring out the analytic jump kernel $J^{B_r(c)}(s,z)$ from the boundary integral. A Walk-on-Spheres sample is cast from $z$, terminating at surface point $t$, which completes the rank-one matrix update: $J^{B_r(c)}(s,z)\cdot d(s,t)\,d(s,t)^\T$. Our utilization of the Beurling-Deny Theorem ensures the matrix update is positive semidefinite, and hence the Monte Carlo estimate robustly recovers the structure of the Steklov eigenspectrum. Pseudocode for this procedure is in Algorithm \ref{alg:interior-psd-estimator}.}
    \label{fig:sampling}
\end{figure}

\paragraph*{Guaranteeing a PSD Estimate.}
To see why the Beurling-Deny decomposition enables a PSD-by-construction estimator, let us again consider a set of boundary basis functions, $\{\phi_k\}_{k=1}^{K}$, which can be evaluated at points $s\in\boundary$. The Steklov stiffness matrix $\stiffness\in\R^{K\times K}$ with entries $\stiffness_{ij}=\E[\phi_i,\phi_j]$ can be expressed via \refeq{eq:beurling-deny-bilinear} as
\begin{align}
    \label{eq:stiffness-matrix-beurling-deny}
    \stiffness = \frac{1}{2}\iint\limits_{\boundary\times\boundary} d(s,t)\,d(s,t)^\T\J^{\vol}(s,t)\ ds\ dt,
\end{align}
where $d(s,t):=b(s)-b(t)\in\R^K$ is the difference vector between the basis evaluations at two boundary points $s$ and $t$. The integrand, $d(s,t)\,d(s,t)^\T \J^\vol(s,t)$, is the product of a rank-one PSD matrix and a non-negative scalar, and hence is itself PSD. What remains is to sample pairs $(s,t)$ according to a distribution related to the jump kernel $\J^\vol(s,t)$. A direct approach to sample the jump kernel is to estimate the differential Poisson kernel $\partial_{n_s}\P^\vol(s\to t)$ using finite-differences, analogous to the derivation of the \hyperref[par:naive-estimator]{na\"ive estimator}.
As discussed in Section \ref{par:naive-estimator}, this approach incurs extreme variance and is brittle when shapes have thin structures.

\paragraph*{Variance Reduction.} Instead of estimating the jump kernel with finite differences, we exploit the fact that the jump kernel is analytically known on a ball (see \refsec{sec:jump-kernel-supp}). By applying the analysis of \citet{yu-2024-tangent-ball}, we are able to write a variance-reduced estimator of \refeq{eq:stiffness-matrix-beurling-deny} by decomposing $\J^\Omega(s,t)$ into two quantities: a closed-form jump kernel for the largest interior ball tangent to $s$, and the result of WoS samples launched from points on the surface of this ball. This decomposition effectively factors out analytically-known jump kernel behavior, casting WoS samples solely for the nontrivial component of the jump kernel that cannot be known ahead of time.

Fix a boundary point $s\in\boundary$ and let $B_r(c)\subset\vol$ be the largest ball contained in $\vol$ that is tangent to $\boundary$ at $s$. The center of this tangent-ball lies at a point $c=s-rn_s$ offset along the inward normal by the ball radius $r$. A Brownian motion starting at a point $x\in B_r(c)$ must first exit the ball at some point $z\in\partial B_r(c)$ before eventually exiting the larger domain $\vol$ at a point $t\in\boundary$. Therefore, we can factor the walk into two independent sub-walks based on these two exit points using the \emph{Strong Markov Property} of Brownian motion,
\begin{align}
    \label{eq:strong-markov-property}
    \P^\Omega(x\to t) &= \int_{\partial B_r(c)} \P^{B_r(c)}(x\to z)\,\P^\vol(z\to t)\ dz.
\end{align}
The first factor,
\begin{align}
    \P^{B_r(c)}(x\to z)&=\frac{1}{4\pi r}\frac{r^2-|x-c|^2}{|x-z|^3},
\end{align}
is the classical Poisson kernel for the ball, known in closed form. The jump kernel for the ball is the normal derivative of its Poisson kernel (\refeq{eq:jump-kernel-poisson}), so for $s\neq z$ we get the analytical representation
\allowdisplaybreaks
\begin{align}
    \label{eq:ball-jump-kernel}
    \J^{B_r(c)}(s,z) &= \frac{1}{2\pi|s-z|^3}.
\end{align}
By writing the jump kernel in terms of the Poisson kernel (\refeq{eq:jump-kernel-poisson}) and using the Strong Markov Property (\refeq{eq:strong-markov-property}), we obtain
\begin{align}
    \label{eq:volume-jump-kernel}
    \begin{aligned}
    \J^\vol(s,t) &= \int_{\partial B_r(c)} \J^{B_r(c)}(s, z)\,\P^\vol(z\to t)\ dz,
    \end{aligned}
\end{align}
proving that we only need to simulate random walks starting from the tangent-ball surface (see Appendix \refsec{sec:jump-kernel-supp} for derivations).

\paragraph{Improved DtN Estimator} Substituting this jump kernel decomposition into the Beurling-Deny representation (\refeq{eq:stiffness-matrix-beurling-deny}) yields
\begin{align*}
    \stiffness &= \frac{1}{2}\iint\limits_{\boundary\times\boundary} d(s,t)\,d(s,t)^\T\left(\int_{\partial B_r(c)} \J^{B_r(c)}(s, z)\,\P^\vol(z\to t)\ dz\right)\ dt\ ds,
\end{align*}
which equivalently rearranges as
\begin{align*}
    \stiffness &= \frac{1}{2}\mkern-16mu\iint\limits_{\mskip24mu\boundary\times \partial B_r(c)}\mkern-8mu\left(\int_{\boundary} d(s,t)\,d(s,t)^\T \,\P^\vol(z\to t)\ dt\right)\,\J^{B_r(c)}(s, z)\ dz\ ds.
\end{align*}
Recognizing the parenthetical as a Poisson integral (\refeq{eq:poisson-kernel-analytic}) allows us to rewrite it as an expectation (\refeq{eq:poisson-kernel-probabilistic})
\begin{align}
    \label{eq:stiffness-matrix-final-integral}
    \stiffness &= \frac{1}{2}\mkern-16mu\iint\limits_{\mskip24mu\boundary\times \partial B_r(c)}\mkern-16mu\Exp_{t}\left[d(s,t)\,d(s,t)^\T\right]\ \J^{B_r(c)}(s, z)\ dz\ ds,
\end{align}
where $t\sim \omega_z^\vol$ is the exit point of a Brownian motion started from $z\in\partial B_r(c)$ on the boundary of the tangent-ball. Finally, we can again write the corresponding Monte Carlo estimator in terms of $s\sim U(\boundary)$ and $z\sim U(\partial B_r(c))$ sampled uniformly.
\begin{empheq}[box=\highlight]{equation}
  \stiffness
    = \ExpUnderIcy{s,\,z,\,t}
    \Biggl[
        \frac{|\boundary|\,|\partial B_r(c)|}{2}\,
        d(s,t)\,d(s,t)^\T\,\J^{B_r(c)}(s, z)
    \Biggr]
\end{empheq}
Algorithm~\ref{alg:interior-psd-estimator} details the pseudocode for our final DtN estimation procedure, which is illustrated in Figure \ref{fig:sampling}.

\begin{prettyalgorithm}[t]{alg:interior-psd-estimator}{Final Interior DtN Estimator}
\begin{algorithmic}
  \Require Boundary mesh $\mathcal{M}$, Galerkin basis $\Phi$, sample count $N$
  \Ensure PSD estimate $\widehat{\*S}\in\R^{K\times K}$ of the DtN stiffness matrix
  \State $\widehat{\*S} \gets \*0 \in \R^{K\times K}$
  \For{$m = 1,\ldots,N$}
  
        \LineComment{Find largest inscribed tangent ball $B_r(c)$ at sampled point $\*s$}
        \State $s, \mathrm{tri}_s \gets $ \Call{UniformSampleMesh}{$\mathcal{M}$}
        \State $r \gets $ \Call{LargestTangentBall}{$s$, $n_s$, $r_{\mathrm{max}}$}
        \State $c \gets s - r\cdot n_s$
        
        \LineComment{Sample random walk starting from tangent ball surface}
        \State $z \gets $ \Call{SampleBallSurface}{$c$, $r$, $n_s$}
        \State $J \gets $ \Call{BallJumpKernel}{$s$, $z$}
        \State $t,\ \mathrm{tri}_t \gets$ \Call{WalkOnSpheres}{$z$}

        \LineComment{Estimate flux of extended basis functions from $\*t$ through $\*s$}
        \State $d \gets \*0 \in \R^{K}$
        \For{$k = 1,\ldots,K$}
            \LineComment{Evaluate $k$-th basis function at $\*s$ and $\*t$}
            \State $b_s \gets $ \Call{BarycentricInterp}{$s,\mathrm{tri}_s,\phi_k$}
            \State $b_t \gets $ \Call{BarycentricInterp}{$t,\mathrm{tri}_t,\phi_k$}
            \LineComment{Store estimated flux of $k$-th basis function}
            \State $d_{\,k} \gets b_s - b_{t}$
        \EndFor
        \LineComment{Update $\widehat{\*S}$ with rank-one PSD sample}
        \State{$\widehat{\*S} \gets \widehat{\*S} + \frac{|\boundary| \cdot |\partial B_r(c)|}{2N} \cdot dd^\T \cdot J$}
  \EndFor
  \State \Return $\widehat{\*S}$
\end{algorithmic}
\end{prettyalgorithm}

\subsection{Exterior DtN Operators}
\begin{figure}
    \centering
    \includegraphics[width=\linewidth , trim=0 0.3cm 0 0]{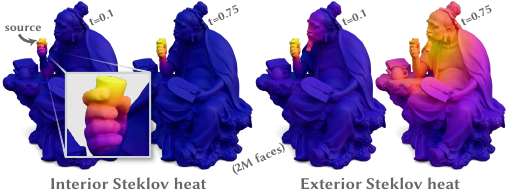}
    \caption{Lu Yu enjoying a hot cup of tea. The exterior Steklov heat warms his face through the ambient volume around the surface. The interior heat evolves slowly by comparison, due to the thin cavity of his hand.}
    \label{fig:lu-yu-heat}
\end{figure}

The DtN operator we have estimated thus far is defined with respect to the \emph{interior} of a given closed boundary. For shapes that are composed of multiple disjoint bounding surfaces---e.g. a mesh of a person holding an apple---these boundaries do not interact, as the Brownian motions simulated in either domain cannot terminate on the opposing surface. To complement our interior DtN operator, we consider its \emph{exterior} variant, defined by an equivalently posed Dirichlet problem on the unbounded exterior domain $\vole:=\R^3\setminus\overline{\vol}$ (i.e., the complement of $\vol$ sharing the same boundary) with the additional Dirichlet condition: $u(x)\to 0$ as $|x|\to\infty$. The resulting operator, denoted $\dtne$, is the \emph{exterior DtN operator}, which treats the unbounded exterior as a single unified volume (see Figure \ref{fig:lu-yu-heat}), allowing it to capture boundary-to-boundary interactions between surfaces that otherwise enclose completely separate shapes.

 \paragraph{Escape to Infinity.} A Brownian motion started in a bounded interior domain $\vol \in \mathbb{R}^3$ is guaranteed to hit the boundary at some point in time. For a Brownian motion in the unbounded exterior domain $\vole$, however, one of two things can happen:

 \medskip
 \begin{enumerate}
    \item it \emph{hits} the boundary and terminates.
    \smallskip
    \item it \emph{escapes} and never returns.
 \end{enumerate}
 
 \medskip\noindent
 In general, it is difficult to Monte Carlo sample first-passage points in the exterior domain because random walk length is unbounded; no matter how far a particle is from the surface, it always has a nonzero probability of returning to the boundary.

\paragraph{Exterior DtN} The Beurling-Deny formula decomposes these two possibilities---that a random walk hits the boundary or escapes---as two components of a single boundary process. Unlike the interior operator, which only has a jump term (\refeq{eq:beurling-deny-bilinear}), the exterior operator is composed of both a jump term and a \emph{killing} term \cite[Eq. 5.8.9]{chen-fukushima-2011-dtn}:
\begin{align}
\label{eq:exterior-dtn-beurling-deny-form}
\begin{aligned}
    \E_{\text{ext}}[f,g] &= \frac{1}{2}\iint\limits_{\boundary\times\boundary} \big(f(s)-f(t)\big)\big(g(s)-g(t)\big)\,\J^{\vole}(s,t)\ ds\ dt\\
    &\quad+\int_\boundary f(s)\,g(s)\,\K^{\vole}(s)\ ds.
\end{aligned}
\raisetag{18pt}
\end{align}
We call $\K^{\vole}:\boundary\to\R$ the \emph{killing measure} of the exterior domain $\vole$ because it models the local rate at which the associated boundary process is killed by escaping to infinity. \citet[Eq. 5.8.6]{chen-fukushima-2011-dtn} define $\K^{\vole}$ in terms of \emph{escape probability},
\begin{align}
    \label{eq:escape-probability}
    q(x) &:= 1-\H^{\vole}[\*1](x).
\end{align}
Then, the killing measure is the normal derivative of this probability,
\begin{align}
    \label{eq:killing-measure-normal-derivative}
    \K^{\vole}(s) &= \partial_{n_s} q(s).
\end{align}
A faithful estimator of $\dtne$ must accurately model both the exterior jump kernel $\J^{\vole}$ and killing measure $\K^{\vole}$, which entails launching random walks into the infinite void. Handling this unboundedness in an unbiased and sample efficient way remains challenging without a shift in perspective.

\paragraph{The Kelvin Transform.} The obstacle to Monte Carlo sampling $J^{\vole}$ and $\K^{\vole}$ is the lack of termination guarantees for random walks in the exterior domain. This issue is rectified using a change of coordinates introduced to the graphics community by \citet{nabizadeh-2021-kelvin}: the \emph{Kelvin transform} is a conformal mapping $\kelvin: \R^3\cup\{\infty\}\to\R^3\cup\{\infty\}$ sending $x\mapsto x/|x|^2$ (with $0\mapsto\infty$ and $\infty\mapsto 0$). Assuming $0\in\vol$ (which can always be arranged by translation), define $\kelvin{\vole}:=\kelvin(\vole\cup\{\infty\})$. Under the Kelvin transform, $\kelvin{\vole}$ is a \emph{bounded} domain in $\R^3$ that includes the origin (as the image of $\infty$).

\paragraph{Sampling Exterior Quantities} The Kelvin transform is exceptionally useful in the harmonic exterior setting. In 3D, a function $u(x)$ (with decay at infinity) is harmonic at $x\in \vole$ if and only if the function $\kelvin[u](\kelvin{x}):=|x|\,u(x)$ is harmonic at $\kelvin{x}\in \kelvin{\vole}$, where $\kelvin{x}:=\kelvin(x)$ denotes the image of $x$ under the Kelvin transform (see Appendix \refsec{sec:kelvin-transform-supp}). Ultimately, this correspondence gives rise to proxies for the exterior domain jump kernel
\begin{align}
    \label{eq:exterior-jump-kernel-final}
    \J^{\vole}(s,t) &= |\kelvin{s}|^3\,|\kelvin{t}|^3\,\J^{\kelvin{\vole}}(\kelvin{s},\kelvin{t})
\end{align}
and the killing measure
\begin{align}
    \label{eq:killing-measure-poisson-kernel}
    \K^{\vole}(s) &= 4\pi\,|\kelvin{s}|^3\,P^{\kelvin{\vole}}(0\to\kelvin{s}),
\end{align}
both of which are written in terms of the bounded domain $\kelvin{\vole}$ where Walk-on-Spheres regains its termination guarantees. The identities in \refeq{eq:exterior-jump-kernel-final} and \refeq{eq:killing-measure-poisson-kernel} are derived in Appendix Sections \ref{sec:exterior-jump-kernel-supp} and \ref{sec:killing-measure-supp} respectively.

\paragraph{An Exterior DtN Estimator} Because $\J^{\kelvin{\vole}}$ is an \emph{interior} jump kernel on a bounded domain, it admits the same tangent-ball decomposition used in \refeq{eq:volume-jump-kernel}. Then, the very same integral rearrangement used in \refsec{sec:improved-beurling-deny-estimator}---this time applied to the exterior Beurling-Deny formula for DtN (\refeq{eq:exterior-dtn-beurling-deny-form})---produces an exterior estimator that operates in the Kelvin-transformed domain.
\begin{boxeq}[eq:main-exterior-estimator][left=3pt][-4ex][-5ex]
\begin{aligned}
    \stiffness_{\text{ext}}
    &= \ExpUnderIcy{s,\,\kelvin{z},\,\kelvin{t}}
    \Biggl[
        \frac{|\boundary|\,|\partial B_r(\kelvin{c})|\,|\kelvin{s}|^3}{2\,|\kelvin{t}|}\,
        d(s,t)\,d(s,t)^\T\,\J^{B_r(\kelvin{c})}(\kelvin{s}, \kelvin{z})
    \Biggr]\\
    &\ \ \ \,+ \ExpUnderIcy{\kelvin{t_0}}
    \Biggl[\frac{4\pi}{|\kelvin{t_0}|}\,b(t_0)\,b(t_0)^{\T}\Biggr]
\end{aligned}
\end{boxeq}
The WoS hit points $\kelvin{t}\sim\omega^{\kelvin{\vole}}_{\kelvin{z}}$ and $\kelvin{t_0}\sim \omega_0^{\kelvin{\vole}}$, for the jump term and killing term respectively, are obtained by casting a Brownian motion from a tangent ball sample $\kelvin{z}\sim U(\partial B_r(\kelvin{c}))$ and $0\in\kelvin{\vole}$ in the inverted domain. Surface points $s\sim U(\boundary)$ are still sampled uniformly on the \emph{primal} boundary. A derivation of the exterior estimator is given in Appendix \refsec{sec:exterior-dtn-estimator-supp}.

\subsection{Boundary function spaces}
\label{sec:function_spaces}
To recover the first $K$ Steklov eigenmodes, we observe that it suffices to work with a discrete representation of the DtN operator that is significantly smaller than the full $V\times V$ dense matrix. We directly estimate a $K\times K$ reduced DtN operator that acts within a projected space $\R^K\subseteq\R^V$. 
Hence, we follow the \emph{Ritz-Galerkin method} to approximate the true Steklov eigenvalues and eigenvectors using an orthonormal basis $\{\phi_1,\phi_2,\ldots,\phi_K\}$ defined on $\boundary$. Because the estimators derived in the preceding sections are written only in terms of evaluations of a finite boundary basis---i.e. $b(s)$ and $d(s,t)$---we consequently need not restrict ourselves to Galerkin bases induced by mesh connectivity (e.g. the ordinary piecewise-linear function space).
Any compact basis on $\partial\Omega$ can be used, provided it can be evaluated at arbitrary points. 

This separation motivates the use of an alternative function space---one that lives on \emph{points} sampled on the surface---that is completely decoupled from the surface discretization itself, which can be poor for in-the-wild geometry. We first discuss our procedure when operating under a mesh-based function space, then we expand on the point-based alternative.

\begin{figure}
    \centering
    \includegraphics[width=\linewidth , trim=0 0.3cm 0 0]{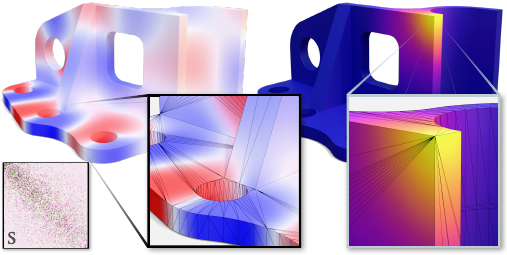}
    \caption{
    Our point-based estimator fully decouples operator resolution from that of the surface discretization, making it robust to exceedingly poor triangulations. By contrast, traditional methods require retriangulation to operate in this regime, which is i) expensive, ii) not robust when the shape is multi-component, and iii) may degrade surface geometry.}
    \label{fig:bad_triangles}
\end{figure}

\paragraph{Mesh-based Galerkin basis.}
Our default choice for the Galerkin basis $\Phi$ is the low-frequency eigenmodes of the cotangent Laplacian defined on the boundary mesh. Basis evaluation, $b(s)$, is performed by barycentric interpolation in each triangle. This choice is efficient and works well even when the boundary mesh has multiple connected components, despite the Laplacian (and its eigenmodes) being decoupled. Our estimated DtN operators nevertheless are able to couple components through the volumetric domain.

\begin{figure}
    \centering
    \includegraphics[width=\linewidth , trim=0 1.8cm 0 0]{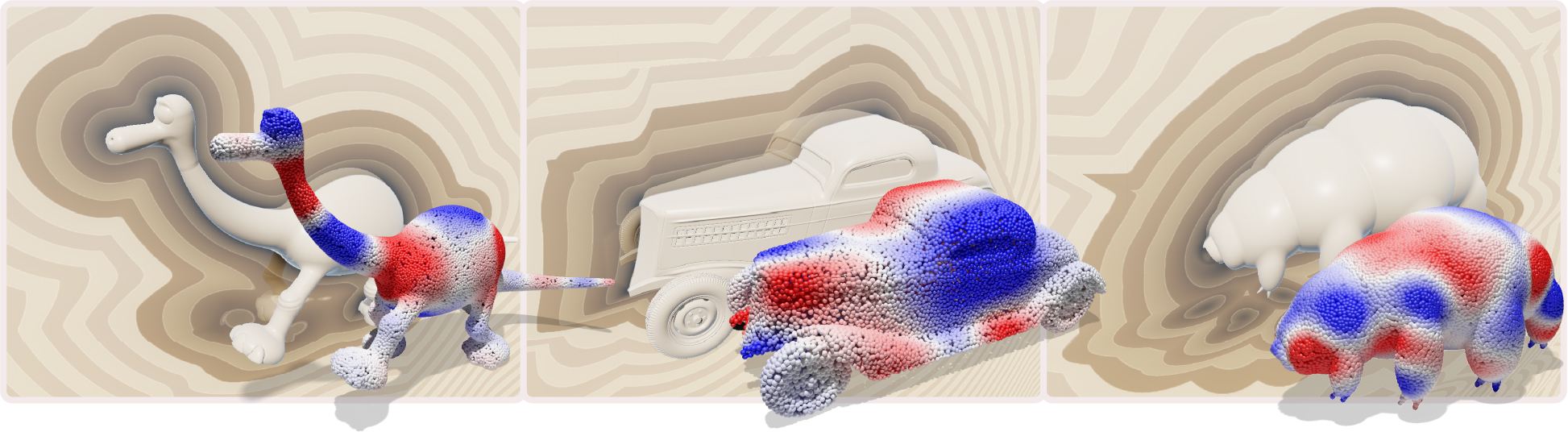}
    \caption{Our point-based DtN estimator can be applied directly to signed distance fields without intermediate meshing. We visualize Steklov eigenfunctions on their respective point cloud function spaces.}
    \label{fig:sdf_steklov}
\end{figure}
\paragraph{Point-based Galerkin basis.}
For meshes with extremely poor surface triangulation, it is in some sense hopeless to represent interesting functions directly in the linear elements themselves (i.e. without retriangulation or similar interventions). The flexibility of Walk-on-Spheres allows us to separate the \emph{geometry} that governs Brownian motions from the \emph{function space} in which we represent our solutions~\cite{sawhney2020mc}. With this fact in mind, we opt to define a function space on a set of densely sampled points on the mesh surface. We sample points $P=\{p_i\}_{i=1}^{N_p}$ on $\partial\Omega$ using blue noise sampling~\cite{bridson2007blueNoise}, and compute the low-frequency eigenmodes of the associated point cloud Laplacian~\cite{robustLaplacian}, which serve as the boundary basis for this representation. During Monte Carlo sampling, WoS paths still terminate on the input mesh, and we interpolate basis values from the sampled points $P$ to the termination point using a $k$-NN interpolator. In particular, we consider a bilateral weighting on both Euclidean distance and normal angular deviation over the $k$ nearest point samples (see Appendix \ref{sec:supp_knn} for details). Replacing barycentric interpolation with this evaluator leaves the estimator unchanged. This scheme can be applied to other surface representations as well, e.g. signed distance fields (see Figure \ref{fig:sdf_steklov}).
\begin{figure}
    \centering
    \includegraphics[width=\linewidth , trim=0 0.3cm 0 0]{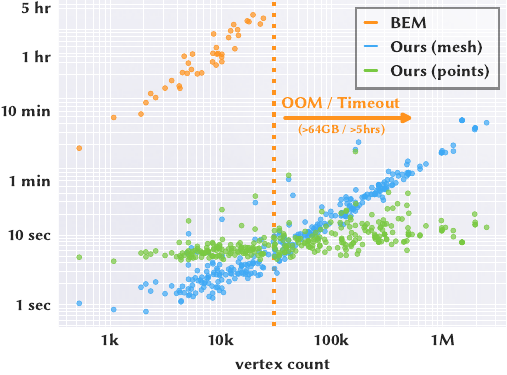}
    \caption{\textbf{Runtime of methods for computing the first 128 interior Steklov eigenmodes.} Our method (10M Monte Carlo samples) is orders of magnitude faster than the BEM-based alternative. Galerkin basis precomputation is included in runtime, making the fixed-size point representation fastest on average--- $2^{14}$ points were used. Out-of-memory (OOM) and runtime constraints precluded running BEM beyond moderate vertex counts.}
    \label{fig:runtime_comparison}
\end{figure}

\begin{figure*}
    \centering
    \includegraphics[width=\linewidth, trim=0 0.3cm 0 0]{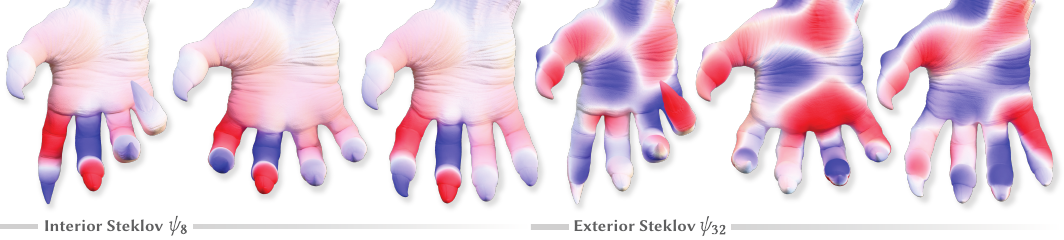}
    \caption{Our interior and exterior Steklov estimator applied to an extremely high-resolution articulating hand with ${\sim}4$ million faces. The interior eigenfunctions are relatively stable under minor deformations; whereas the exterior counterparts are more sensitive to relative positions of the fingers and palm.} \label{fig:creature_hand_eigvecs}
\end{figure*}

%% file: content/sections/implementation.tex
\section{Implementation}
\label{sec:implementation}
\interfootnotelinepenalty=10000

We implement our estimators in CUDA, using \verb|cuBQL| BVH \cite{cuBQL} as our only major dependency for WoS distance queries. In the exterior domain, we model inverted triangle meshes with spherical-triangular elements and employ appropriate (non-planar) distance queries to these elements. Unless otherwise specified, we forgo sampling of the killing measure---effectively fixing it to zero---which allows the exterior operator to behave conservatively\footnote{The killing measure is the only Beurling-Deny term that acts on constants. Dropping it restores a constant zero eigenmode, making the associated heat flow preserve constants instead of dissipating functions to zero, as is typical for diffusion operators used in geometry processing.}. We use a WoS termination tolerance of \verb|1e-6| for all experiments. Tangent balls are computed using the bisection algorithm of \citet{yu-2024-tangent-ball} with 10 iterations; similarly, we perform antithetical sampling on each tangent ball. We importance sample tangent balls according to their jump kernels and forgo sampling points on the ball especially close to the surface point $s$ to avoid the singularity. Namely, we exclude a spherical cap subtending $\sim\!2.5^\circ$ about $s$ from the domain of integration. Our CUDA kernels sort sample points $s\in\boundary$ via Z-order curves, which greatly reduces warp divergence; additionally, the outer product $d(s,t)\,d(s,t)^T$ is deferred to be computed in batches to avoid shared memory contention of the empirical operator $\*S$. Finally, we note that our theoretical treatment in Section \ref{sec:monte-carlo} is with respect to much stricter conditions (i.e. smooth, closed domains) than the applied setting we proceed with---we refer to Appendix \ref{sec:supp_input_assumptions} for clarifying details surrounding this practical gap.

%% file: content/sections/evaluation.tex
\begin{figure}
    \centering
    \includegraphics[width=\linewidth , trim=0 0.3cm 0 0]{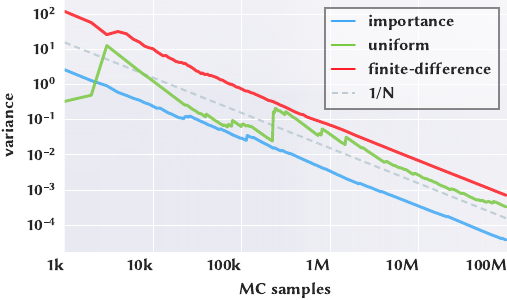}
    \caption{\textbf{Mean matrix-entry variance of our estimated DtN operator.} Our estimators converge under the usual $O(1/N)$ variance w.r.t. the number of Monte Carlo samples. The finite-difference estimator (red) exhibits highest variance across all sample counts. Employing the integral decomposition in Section \ref{sec:improved-beurling-deny-estimator} and uniformly sampling the surface of tangent balls reduces variance (green). Importance sampling the jump kernel on these balls further reduces variance and stabilizes the estimator (blue).}
    \label{fig:variance_comparison}
\end{figure}

\begin{figure*}
    \centering
    \includegraphics[width=\linewidth, trim=0 0.4cm 0 0]{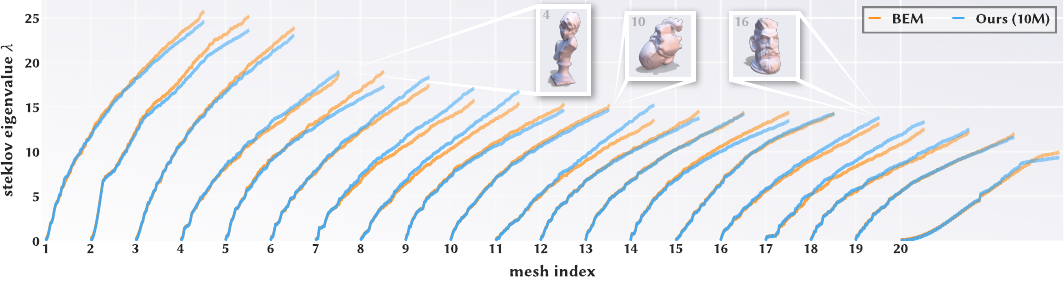}
    \caption{Our estimated interior Steklov eigenspectra computed on 20 preprocessed shapes from Thingi10k, compared to the BEM-based method of \citet{wang-2017-steklov}. Each pair of eigenspectra represent a unique shape. Our method faithfully approximates the Steklov eigenmodes while reducing runtime massively. We note that BEM employs an iterative method that admits error, hence exact replication---especially near the tail-end of the spectrum---is not expected.} \label{fig:thingi_bem_comparison}
\end{figure*}

\begin{figure*}
    \centering
    \includegraphics[width=\linewidth, trim=0 0.25cm 0 0]{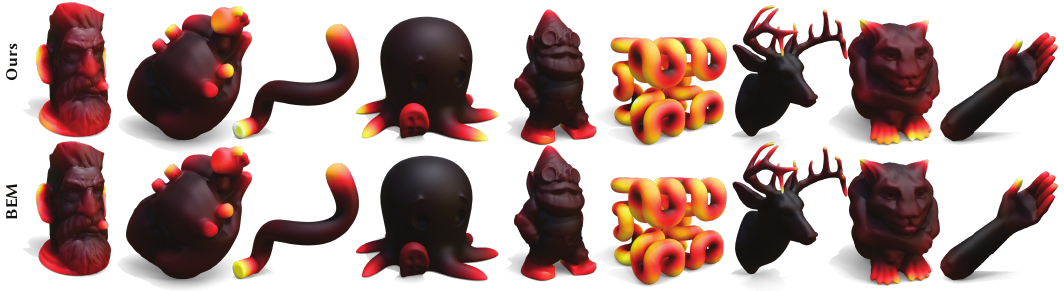}
    \caption{Comparison of heat kernel signatures derived from the interior Steklov eigenspectra, as computed by our Monte Carlo estimator and BEM.} \label{fig:hks_comparison}
\end{figure*}

\section{Evaluation}
\label{sec:evaluation}
We compare our Monte Carlo estimators to the method of \citet{wang-2017-steklov} (denoted \verb|BEM|), which employs a matrix-free boundary element method to resolve interior Steklov eigenmodes. We first demonstrate that our interior DtN estimator yields a matrix whose eigenspectrum faithfully converges to that of BEM. We then explore the unique scalability and robustness of our method, showing that it can be rapidly applied to large meshes and ones with complex or degenerate surface discretizations.

\paragraph{Setup.} We apply the open source implementation of \citet{wang-2017-steklov} to compute the first $k\!=\!128$ interior Steklov eigenmodes on shapes sourced from Thingi10k~\cite{zhou2016thingi10k}, using 25 LOBPCG iterations. Data that is labeled as \verb|preprocessed| implies that the corresponding shapes have undergone retriangulation using fTetWild \cite{hu2020ftetwild} surface extraction---we found this necessary to ensure BEM converges on shapes with poor discretizations. We employ our Monte Carlo estimator on the same shapes, using a Galerkin basis proportionally sized to $k$---in practice, we use $k+32$ basis functions. We evaluate our mesh-based and point-based variants as defined in Section \ref{sec:function_spaces}. Reported runtimes for our method are recorded w.r.t. a consumer-grade NVIDIA RTX 4090.

\paragraph{Convergence.} We establish the correctness of our estimator by directly comparing its eigenmodes to BEM on preprocessed Thingi10k shapes. In particular, Figure \ref{fig:thingi_bem_comparison} compares the eigenvalues of both methods across 20 shapes. Comparing the \emph{eigenvectors} themselves is less trivial due to sign and permutation ambiguity; further, repeated eigenvalues are valid under arbitrary orthogonal rotations. For this reason, we opt to compare eigenvectors through basis-invariant quantities induced by the corresponding spectral decompositions. In particular, Figure \ref{fig:hks_comparison} compares the Heat Kernel Signatures (HKS) computed using both spectra, showing that---up to subtle numerical differences---their structure is nearly identical. Finally, Figure \ref{fig:variance_comparison} verifies that our estimators converge under $O(1/N)$ variance in the number of samples, as expected of Monte Carlo methods; the integral decomposition of Section \ref{sec:improved-beurling-deny-estimator} specifically vastly reduces variance as compared to the finite-difference baseline.

\paragraph{Robustness.} Our DtN estimators inherit the robustness of Monte Carlo PDE solvers, making them directly applicable to shapes with hundreds of connected components, degenerate triangulations, and self-intersections. In Figure \ref{fig:bad_triangles} we show that our construction is directly applicable to a CAD-like model that is primarily composed of sliver triangles; in spite of this, the estimated Steklov modes are meaningful and produce correct heat-like behavior. Similarly, Figure \ref{fig:multi_component} shows the behavior of our interior and exterior Steklov modes on a non-trivial \texttt{Commodore 64} mesh that contains 2662 connected components with poor element quality. The interior Steklov heat dissipates beneath the keycaps, diffusing slowly due to the disproportionate number of obstructions, compared to the exterior heat, which diffuses over a broader profile. We define \emph{Steklov heat} as the heat equation associated with the DtN operator~\cite{wang-2017-steklov},
\begin{align}
    \frac{du}{dt} = -\Lambda u,
    \label{eq:steklovHeat}
\end{align}
whose solution is approximated in the Steklov eigenbasis as
\begin{align}
    u_t = \sum_{k=1}^K e^{-\lambda_k t} a_k \psi_k,
    \label{eq:spectralSteklovHeat}
\end{align}
where $(\lambda_k, \psi_k)$ are the $k$-th Steklov eigenpair, and $a_k = \langle u_0, \psi_k\rangle_{\*M}$ represents spectral coefficients of the initial heat distribution. The DtN operator, $\Lambda$, can be taken as either the interior or exterior variant. To visualize the fundamental behavior of Steklov heat, Figures \ref{fig:multi_component}, \ref{fig:bad_triangles}, and \ref{fig:lu-yu-heat} take their initial heat distributions, $u_0$, as a unit impulse localized at one vertex. Although $\Lambda$ and its eigenmodes live on the boundary, the Steklov heat equation behaves volumetrically, and dissipates \emph{through} the volume implied by a surface mesh.

\paragraph{Scalability.} The CUDA implementation of our estimators is able to use modern GPUs effectively, making it substantially more efficient than traditional volumetric spectral methods, which are ordinarily CPU-bound. Figure \ref{fig:runtime_comparison} compares the runtime of our estimator at 10 million samples against that of BEM. We find that our method is orders of magnitude faster across all mesh sizes, and is able to handle much larger element counts, whereas BEM eventually fails due to out-of-memory errors or compute timeout (5 hours). Figure \ref{fig:objaverse_timings} shows timings for our interior and exterior estimators extended to the large-scale, uncurated Objaverse dataset~\cite{deitke2023objaverse}. We discuss further details and findings for this experiment in Section \ref{sec:representationLearning}---critically, scaling existing tetrahedral or BEM-based methods to the size and topological irregularity of this dataset is infeasible.

%% file: content/sections/applications.tex
\begin{figure*}[p]
    \centering
    \includegraphics[width=\linewidth, trim=0 0.2cm 0 0]{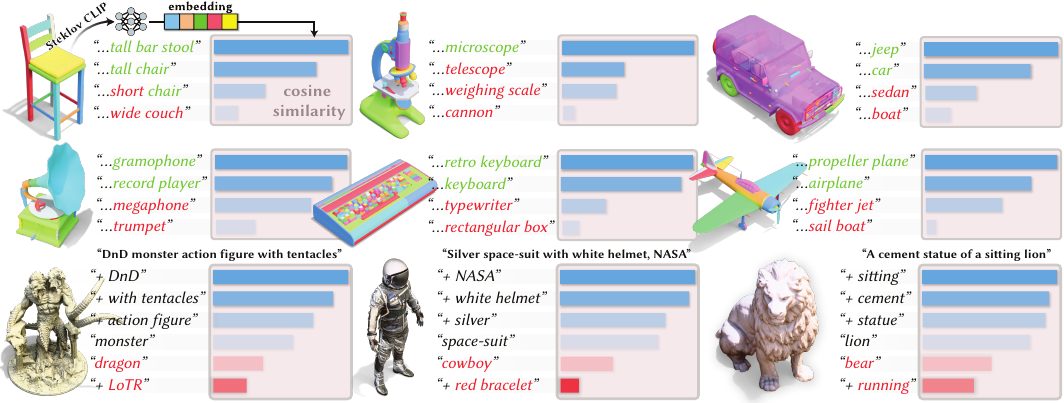}
    \caption{\textbf{Qualitative results from our Steklov-CLIP model.} We probe representations learned by Steklov-CLIP on a range of meshes sourced from public repositories---all meshes shown have one or more of the following qualities: many connected components, poor element quality, or extremely dense triangulation (i.e. representative qualities of in-the-wild shapes). We probe our model by comparing cosine similarities of its shape embeddings to manually authored text queries. Green and red words indicate terms that are semantically relevant or dissimilar to the objects, respectively. The bottom row shows cosine similarity changing as more query terms are included. Similarity scores depicted as colored bars are drawn relative to the highest attained score.} \label{fig:clip_quantitative}
\end{figure*}

\begin{figure*}[p]
    \centering
    \includegraphics[width=\linewidth, trim=0 0.3cm 0 0]{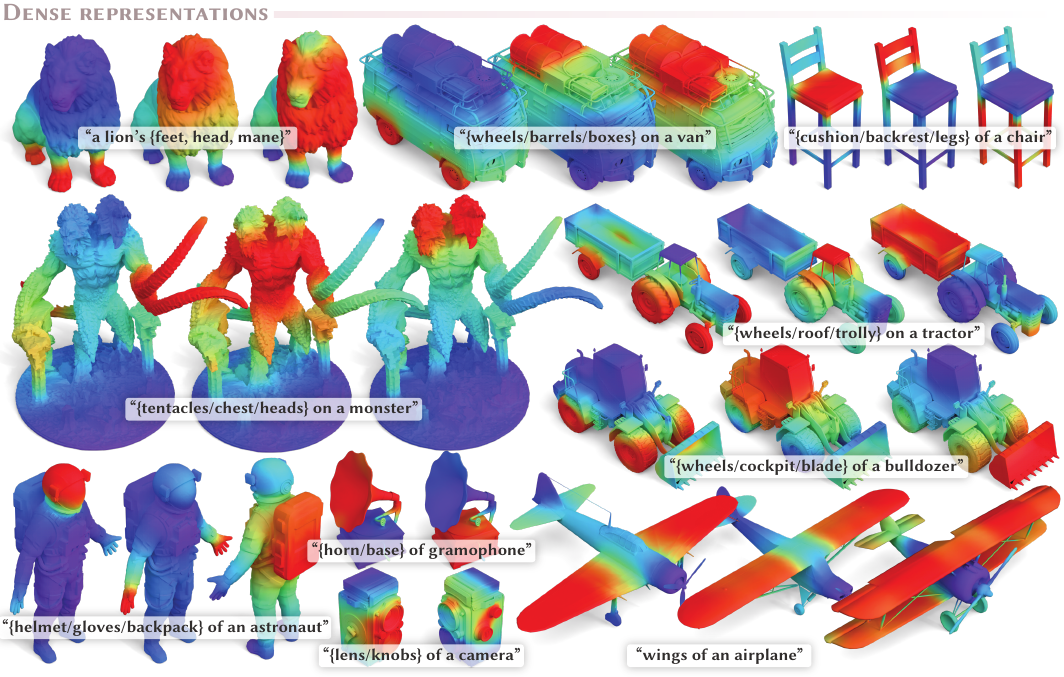}
    \caption{\textbf{Saliency maps from our finetuned Steklov-CLIP.} We visualize cosine similarity of per-point embeddings produced by our model against text embeddings (overlaid). Our finetuned model is able to localize semantic parts, and even generalizes to heavy-tail cases (e.g. the two-headed Demogorgon).} \label{fig:clip_saliency}
\end{figure*}

\section{Representation Learning on Meshes}
\label{sec:representationLearning}
The preceding sections make interior and exterior Steklov eigenmodes available at the scale and irregularity of modern shape datasets. To demonstrate this fact further, and to highlight the utility of the Steklov modes themselves, we employ these spectra as primitive building blocks in a representation learning pipeline for meshes. In particular, we use Steklov operators to define volumetrically-informed feature transformations in a neural architecture, in similar style to previous (intrinsic) networks for meshes~\cite{smirnov2021hodgenet,sharp2022diffusionnet,maesumi2025poissonnet}. 

Learning 3D shape representations with contrastive pre-training has been studied in several works, which can largely be categorized as point cloud-based networks, and ones that use multi-view images---we refer to~\citet{lee2025duoduo} for an overview. To the best of our knowledge, there has yet to be a faithful attempt at large-scale contrastive pretraining on meshes directly, simply because the necessary geometry processing machinery does not exist---the limitations of intrinsic methods preclude their application to the kinds of datasets at hand, and traditional volumetric methods are too expensive to be applied at this scale. Our goal, then, is to train a mesh encoder whose embeddings lie in the same semantic space as a frozen text-image contrastive model (e.g. CLIP).

\paragraph{Preliminaries}
Following the standard recipe in this domain, we consider a large collection of shapes obtained from the Objaverse dataset~\cite{deitke2023objaverse}, from which we have corresponding caption and multi-view image embeddings. In particular we use the embeddings provided by \citet{lee2025duoduo}, which employs the \verb|laion2b_s34b_b79k| OpenCLIP checkpoint~\cite{ilharco2021openclip}. We precompute interior and exterior Steklov eigenspectra on $\sim\!450,\!000$ shapes as outlined in Section \ref{sec:evaluation}.
We discard $\sim\!30,\!000$ shapes that were of extremely low quality.
Our network is trained using an InfoNCE contrastive objective~\cite{oord2018infoNCE} defined over paired batches of embeddings from two modalities. Let $E_a = \{e^a_i\}_{i=1}^N$ and $E_b = \{e^b_i\}_{i=1}^N$, where $e^a_i,e^b_i \in \mathbb{R}^d$ denote embeddings of the $i$-th paired shape instance in modalities $a$ and $b$. Then the one-way contrastive loss is
\begin{align}
    \ell_{a \to b}(E_a,E_b)
    =
    -\frac{1}{N}\sum_{i=1}^N
    \log
    \frac{
        \exp\left(\langle e^a_i, e^b_i\rangle / \tau\right)
    }{
        \sum_{k=1}^N
        \exp\left(\langle e^a_i, e^b_k\rangle / \tau\right)
    },
    \label{eq:infoNCE}
\end{align}
where $\tau$ is a learned temperature. We consider three modalities, yielding embeddings for shape $S$, text $T$, and images $I$, making the final symmetric objective
\begin{align}
    \mathcal{L}_{\mathrm{CLIP}}
    =
    \tfrac{1}{4}
    \left(
        \ell_{S \to T} + \ell_{T \to S} + \ell_{S \to I} + \ell_{I \to S}
    \right).
\end{align}
\begin{figure}
    \centering
    \includegraphics[width=\linewidth , trim=0 0.3cm 0 0]{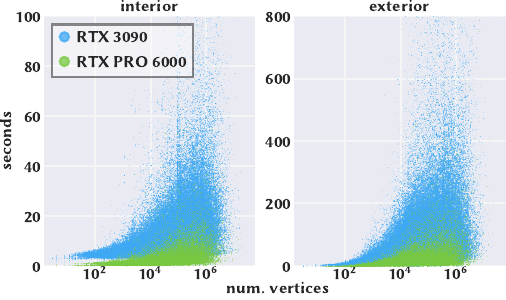}
    \caption{Compute timings of our interior and exterior estimators (10M Monte Carlo samples) applied at scale to $\sim\!450$k shapes from Objaverse. Compute was distributed across a mix of RTX 3090 and RTX PRO 6000 GPUs. Our method effectively utilizes modern GPUs and hence we observe vastly better throughput on the latter (more modern) Blackwell chips. Further, the exterior estimator is generally slower due to Kelvin-inverted BVH distance queries, which are generally more expensive than ordinary triangular mesh queries.}
    \label{fig:objaverse_timings}
\end{figure}

\paragraph{Network Architecture}
Our network architecture takes inspiration from DiffusionNet \cite{sharp2022diffusionnet} and Galerkin Transformer \cite{cao2021galerkinTransformer}. More specifically, the former defines feature transformations through an intrinsic heat equation, solved via a spectral approximation---which takes the same form as \refeq{eq:spectralSteklovHeat}---defined by the eigenspectrum of the Laplace-Beltrami operator. Our network employs a similar philosophy, though it uses interior and exterior Steklov spectra in place of this intrinsic alternative, which endows our network with the ability to propagate features through volumetric regions that are possibly disconnected. In particular, our core network block begins by applying Steklov heat filters to the incoming feature field, denoted $\*f: \boundary\to\R^C$. Letting $\stekBasisInt$, $\stekBasisExt$ denote the interior and exterior Steklov bases, we apply per-channel
\begin{align}
    \*h^c_{\mathrm{s}} \gets \*\Psi_s \begin{bmatrix}e^{-\lambda^0_{\mathrm{s}} t_{\mathrm{s}}^c}\\... \\e^{-\lambda^k_{\mathrm{s}} t_{\mathrm{s}}^c}\end{bmatrix}\odot (\*\Psi_s ^\T \*M\,\*f_{\mathrm{s}}^c), \quad s\in\{\mathrm{int},\mathrm{ext}\}.
\end{align}
For brevity we use $\cdot_s$ when applying operations with both Steklov spectra, and $\*t_s\in\R^{C/2}$ is a learned per-channel heat time. Interior and exterior heat filters are applied to both halves of the feature vector, and the resulting features are concatenated back together.

Galerkin Transformer motivates a second ingredient in our architecture, though in a more indirect way. The relevance to our setting is in its observation that a softmax-free attention layer can be viewed as an operator assembled from inner products of functions, rather than a dense pair-wise interaction between tokens---or in our case, quadrature points. 
We employ this idea using Steklov eigenspaces as the underlying function space. In particular, we first apply a learned linear projection and then represent the resulting feature fields in the Steklov spectral domains,
\begin{align}
     \*u = \*f\,\*W_{\mathrm{in}}, \qquad
    \*Z_{\mathrm{int}} = \stekBasisInt^\T \*M \*u, \quad
    \*Z_{\mathrm{ext}} = \stekBasisExt^\T \*M \*u .
\end{align}
A learned linear transformation then produces three coefficient fields, which represent query, key, and value matrices whose columns are scalar boundary functions represented in the Steklov spectrum
\begin{align}
    \*Q_{s} &= \*Z_s \*W^{Q}_{s}, &
    \*K_{s} &= \*Z_s \*W^{K}_{s}, &
    \*V_{s} &= \*Z_s \*W^{V}_{s},
    \quad
    s\in\{\mathrm{int},\mathrm{ext}\}.
\end{align}
We additionally modulate these functions by learned eigenvalue-dependent filters, in particular learned mixtures of heat kernels given by \refeq{eq:spectralSteklovHeat}.
Our block then forms the bilinear operators
\begin{align}
    \*G_{\mathrm{int}}
    =
    \tfrac{1}{k}
    \*K_{\mathrm{int}}^\T \*V_{\mathrm{int}},
    \qquad
    \*G_{\mathrm{ext}}
    =
    \tfrac{1}{k}
    \*K_{\mathrm{ext}}^\T \*V_{\mathrm{ext}},
\end{align}
where $k$ is the size of our Steklov basis. In classical attention, this intermediate matrix (taken instead between queries and keys) would represent dense similarity between all quadrature points; here, however, $\*G$ is a small operator on feature channels assembled by integrating over the Steklov function space, and notably its size is independent of the number of quadrature points on the surface. 

We allow $\*G_{\mathrm{int}}$ and $\*G_{\mathrm{ext}}$ to exchange information through the learned update
\begin{align}
    \begin{bmatrix}
        \*G_{\mathrm{int}} \\
        \*G_{\mathrm{ext}}
    \end{bmatrix}
    \leftarrow
    \begin{bmatrix}
        \alpha_{\mathrm{ii}} & \alpha_{\mathrm{ie}} \\
        \alpha_{\mathrm{ei}} & \alpha_{\mathrm{ee}}
    \end{bmatrix}
    \begin{bmatrix}
        \*G_{\mathrm{int}} \\
        \*G_{\mathrm{ext}}
    \end{bmatrix},
\end{align}
where the coupling is learned independently for each attention head and initialized near the identity. Output coefficients are obtained by applying these operators to query vectors
\begin{align}
    \*Y_{\mathrm{int}}
    =
    \tfrac{1}{\sqrt{d}}\,
    \*Q_{\mathrm{int}}\*G_{\mathrm{int}},
    \qquad
    \*Y_{\mathrm{ext}}
    =
    \tfrac{1}{\sqrt{d}}\,
    \*Q_{\mathrm{ext}}\*G_{\mathrm{ext}},
\end{align}
where $d$ is the per-head channel dimension. Finally, we map back to the spatial domain via $\stekBasisInt \*Y_{\mathrm{int}}$, $\stekBasisExt \*Y_{\mathrm{ext}}$. Importantly, all transformations in this block are invariant to the arbitrary choice of eigenbasis (i.e. its symmetries) discussed in Section \ref{sec:evaluation}.

To recapitulate, each network block contains two spectral operations: Steklov heat, which applies a volumetric diffusion-like filter on boundary features, and the Steklov-Galerkin transform above, which learns a data-dependent bilinear interaction between functions represented in the Steklov function spaces. Finally, resulting features from these operations are mixed with an MLP. We summarize the core network block in Algorithm \ref{alg:steklov-net-block}.

\begin{figure}
    \centering
    \includegraphics[width=\linewidth , trim=0 0.3cm 0 0]{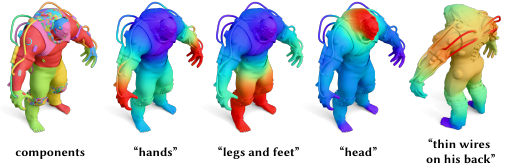}
    \caption{Steklov-CLIP's semantic saliency maps on a mesh with 80 connected components. Our model is even able to discern thin features (right).}
    \label{fig:clip_mutant_saliency}
\end{figure}
\paragraph{Training}
We train our Steklov-CLIP model for \num{45000} iterations on 6$\times$B200 GPUs, using a total batch size of 960. We use AdamW with $\beta_1=0.9,\beta_2=0.99$ and a cosine annealing learning rate schedule with warmup, taking the initial learning rate as \verb|1e-7| scaling to \verb|6e-4| over 500 steps, then tapering to \verb|3e-4| over \num{4000} steps. We use weight decay of $0.01$, which is not applied to any rank-1 trainable parameters (e.g. layer scales and learned heat times). Our network is trained with \texttt{BFloat16} precision and the learned InfoNCE temperature, $\tau$, used in \refeq{eq:infoNCE} is capped to 50. Finally, our network uses xyz coordinates, normals, and RGB colors as input.

\paragraph{Evaluation}
We evaluate Steklov-CLIP on the canonical zero-shot Objaverse-LVIS classification benchmark. Zero-shot classification is performed through cosine similarity of shape embeddings (as predicted by the network) against text embeddings derived from LVIS category labels. In particular, the category labels are augmented through several text templates, e.g. \texttt{a detailed 3D model of a \{category\}}, whose associated embeddings are averaged to form a final target embedding. LVIS contains \num{46207} shapes across \num{1156} categories. In Table \ref{tab:clipEval}, we report top-1 and top-5 zero-shot classification accuracies of our method against representative point cloud-based and multi-view methods. Figure \ref{fig:clip_quantitative} shows a qualitative probe of our model's representations, demonstrating that Steklov-CLIP is able to correctly understand even subtle aspects of input meshes. We further evaluate fine-grained representations learned by these models by performing a dense classification task over all points on the shape. In particular, we use part-level captions from the Partverse~\cite{dong2025partverse} dataset to perform per-point retrieval over all part captions associated with each shape. Table \ref{tab:partverseEval} shows retrieval accuracies in this setting, which are greatly improved by the fine-tuning strategy discussed below.

\begin{table}
    \centering
    \caption{Zero-shot classification scores on Objaverse-LVIS. Shape embeddings are evaluated by nearest-neighbor classification against augmented LVIS category labels. We compare to representative point cloud (PC) and multi-view (MV) methods. $\dagger$ Steklov-CLIP acts on a point-based functional basis (see Sec. \ref{sec:function_spaces}) but uses our mesh-aware Steklov spectra.}
    \label{tab:clipEval}
    \rowcolors{2}{seabornblue!75}{seabornblue!25}
    \begin{tabular}{l | c c | c c}
        \toprule
        \multicolumn{5}{c}{\textbf{Objaverse LVIS Zero-shot Classification}} \\
        \midrule
        \rowcolor{white}
        \textbf{Method}
            & \thead{Rep}
            & \thead{Params}
            & \thead{Top1} 
            & \thead{Top5} \\
        \midrule
        % \midrule
        ULIP~\shortcite{xue2023ulip}
            & PC & 151M & 26.8 & 52.6 \\
        OpenShape~\shortcite{liu2023openshape}
            & PC & 32M & 46.8 & 77.0 \\
        Uni3D-base~\shortcite{zhou2024uni3d}
            & PC & 88M & 51.7 & 80.8 \\
        Uni3D-giant~\shortcite{zhou2024uni3d}
            & PC & 1B & 55.3 & 82.9 \\
        ViT-Lens~\shortcite{lei2024vitlens}
            & PC & 233M & 52.0 & 79.9 \\
        \midrule
        CLIP~\shortcite{radford2021CLIP}
            & MV & 87M & 35.7 & 62.1 \\
        Duoduo CLIP~\shortcite{lee2025duoduo}
            & MV & 87M & 55.2 & 83.4 \\
        \midrule
        \textbf{Steklov-CLIP} \textit{(ours)}
            & Mesh$^\dagger$ & 53M & 49.1 & 76.4  \\
        \midrule
        \addlinespace[0.25em]
    \end{tabular}
\end{table}

\begin{table}
    \centering
    \caption{We evaluate fine-grained semantic alignment by retrieving part captions using per-point shape embeddings. The \texttt{Point} metric classifies each surface point by its nearest part caption embedding. The \texttt{Part} metric first pools per-point embeddings over each ground-truth part before retrieval. \textit{FT} denotes our finetuned model, which is evaluated on held-out data.}
    \label{tab:partverseEval}
    \rowcolors{2}{seabornblue!75}{seabornblue!25}
    \begin{tabular}{l | c c}
        \toprule
        \multicolumn{3}{c}{\textbf{Partverse Part-Caption Retrieval}} \\
        \midrule
        \rowcolor{white}
        \textbf{Method}
            & \thead{Part (T1/T5)} 
            & \thead{Point (T1/T5)} \\
        \midrule
        OpenShape~\shortcite{liu2023openshape}
            & 32.1 / 80.8 & 23.9 / 75.4 \\
        Uni3D-base~\shortcite{zhou2024uni3d}
            & 35.8 / 82.3 & 32.8 / 81.0 \\
        \textbf{Steklov-CLIP}
            & 43.4 / 85.7 & 35.9 / 82.2 \\
        \textbf{Steklov-CLIP} \textit{(FT)}
            & 60.1 / 93.0  & 54.3 / 91.8 \\
        \midrule
        \addlinespace[0.25em]
    \end{tabular}
\end{table}

\paragraph{Fine-grained semantic queries}
CLIP-style contrastive pre-training is a proven tool for high-level semantic understanding; however, previous literature has demonstrated its limitations in extracting meaningful fine-grained representations, i.e. ones that can semantically localize parts of a subject~\cite{zhong2022regionclip,mukhoti2023clipSeg,tschannen2025siglip2}. To enhance the fine-grained capability of our Steklov-CLIP model, we finetune it on a small collection of shapes with part-level captions taken from the Partverse dataset~\cite{dong2025partverse}. We finetune on just \num{6995} shapes and reserve \num{1230} for validation. Each shape has a variable number of parts with corresponding captions, we embed the captions using the same OpenCLIP checkpoint as in our pre-training. We finetune using an objective that encourages both part-level and per-point semantic alignment to these part captions---in particular, we use
\begin{align}
\begin{aligned}
    \mathcal{L}_{\mathrm{finetune}} =  \mathcal{L}_{\mathrm{part}} + \mathcal{L}_{\mathrm{point}} + \gamma\mathcal{L}_{\mathrm{reg}}
\end{aligned}
\end{align}
where $\mathcal{L}_{\mathrm{part}}$ computes InfoNCE between pooled point embeddings on each part; similarly, $\mathcal{L}_{\mathrm{point}}$ contrasts all point embeddings to each part directly, and $\mathcal{L}_{\mathrm{reg}}$ is a low-weight cosine similarity between part shape/text embeddings, taking $\gamma=0.1$. We finetune for only \num{3500} steps using a batch size of \num{720} and learning rate of \texttt{5e-6}.

Even this modest amount of finetuning greatly improves Steklov-CLIP's ability to localize semantic geometric features. Figures \ref{fig:clip_saliency} and \ref{fig:clip_mutant_saliency} show qualitative results in which Steklov-CLIP's point-wise cosine similarity is visualized w.r.t. a given text query. These saliency maps are filtered through a short-time Steklov heat equation to remove noise---see Figure \ref{fig:clip_saliency_unfiltered} for raw saliency maps. Our model is able to isolate semantic parts, and even generalizes to unusual cases---e.g. the two-headed Demogorgon with tentacles.

\paragraph{Remarks}
The preceding evaluations suggest that Steklov-CLIP is perhaps more parameter and data efficient as compared to existing methods, given that our model is trained only on ${\sim}50\%$ of the data as cited works. Further, our model's increased performance on fine-grained evaluation (Table \ref{tab:partverseEval}) likely benefits from: i) not requiring tokenization---unlike the PointBERT transformers used in cited point cloud methods; and b) our use of geometric operators that directly encode local effects. Finally, several methods initialize directly from pretrained weights of internet-scale vision models (i.e. Duoduo CLIP, Uni3D, ViT-Lens), and hence direct comparison should be qualified---this characteristic may also affect the aforementioned models' abilities to localize features, if for example they are too reliant on non-geometric features that are acquired from initialization.

%% file: content/sections/conclusion.tex
\section{Conclusion}
\label{sec:conclusion}

We have demonstrated a practical algorithm for volumetric spectral geometry processing using the Dirichlet-to-Neumann operator. The scalability and robustness of our Monte Carlo method unlocks the utility of these geometric constructs, and has allowed us to apply volumetric techniques to large-scale mesh representation learning, which previously was only viable through point cloud-based and multi-view methods.

\paragraph{Discussion \& Limitations}
Our method is not without limitations---most apparent is that our method employs a spectral approximation to the estimated DtN operators. For many practical applications in geometry processing, spectral approximations are welcomed, though we recognize potential applications in applied sciences that may require higher frequency components in the approximated operators. Further, resolving the \emph{highest}-frequency eigenmodes of such operators may be entirely impractical with Monte Carlo, due to the extreme variations and numerical sensitivity at such scales. 

While our CUDA implementation of the estimators in Section \ref{sec:monte-carlo} is extremely fast, there remain several axes to improve performance. First, as noted by Figure \ref{fig:objaverse_timings}, our exterior DtN estimator is bottlenecked by closest-point queries to the Kelvin-inverted surface, which are more computationally expensive than ordinary point-triangle queries. We suspect more optimizations can be made to the acceleration structures employed, as well as the CUDA code in general (i.e. by reducing overhead in the exterior estimator's kernels). Second, the Galerkin bases discussed in Section \ref{sec:function_spaces} follow from sparse matrices and hence rely on CPU-based eigendecomposition routines. While these bases are effective for our applications, we foresee opportunity to use simpler (and cheaper) options---e.g. Random Fourier Features (RFF) derived directly from vertex coordinates~\cite{rahimi2007RFF}. Especially for large meshes, we are primarily bottlenecked by Galerkin basis computation rather than the Monte Carlo estimator itself. Finally, while our estimators are well-defined in open domains (so long as conventional surface normals are determined), its runtime suffers due to WoS samples taking longer to converge and requiring theoretically unbounded steps. We suspect the application domains at hand would benefit from methods that reduce this burden by introducing bias---much to the dismay of Monte Carlo puritans.

We are excited by the prospect of expanding this methodology to more volumetric operators beyond DtN. For instance, even the Poisson kernel itself immediately yields a spectral decomposition of interest~\cite{auchmuty2017svdPoisson}, to which our method can be directly applied. Second, we hope to see more geometric methods built upon the exterior harmonic processes explored in this paper, e.g. for further downstream applications to multi-component geometry.

%% file: content/sections/supplemental.tex
\newpage
\section{Practical input assumptions}
\label{sec:supp_input_assumptions}

The derivations in Section \ref{sec:monte-carlo} are written with respect to a smooth boundary $\boundary$ of a volumetric domain $\Omega$. In practice, however, our implementation is applied to meshes from minimally-curated shape collections, where surfaces may be open, inconsistently oriented, multi-component, or locally degenerate. This section describes the practical convention used to turn such inputs into two-sided surfaces on which the interior and exterior estimators can be defined.

For watertight consistently oriented meshes, the convention agrees with the usual notion of inside and outside. For partially open surfaces or those with inconsistent normals, we instead use the generalized winding number (GWN) of \citet{jacobson2013winding} to define a canonical outward direction. This construction provides a consistent assignment of sidedness for interior and exterior WoS samples. To summarize:
\medskip
\begin{enumerate}
    \item \textbf{Do we require closed surfaces?} No, we do not require input meshes to be watertight. Normals are used to determine the direction in which interior and exterior tangent balls are drawn, from which walks are sampled. When the input orientation is inconsistent, we repair normals using generalized winding numbers, orienting each face according to the local inside-outside convention. In extreme cases (e.g. a plane) the sides of a surface become geometrically symmetric; in this setting, our interior and exterior operators should be interpreted as the two indistinguishable halves of a ``two-sided'' operator, rather than as distinct operators associated with inside and outside.

    \medskip
    \item \textbf{What about self-intersections?} The largest inscribed tangent ball is not well-defined arbitrarily close to an intersection point; in the limit, its radius approaches zero. In practice, we do not require the mesh to be intersection-free. Samples near these singularities are discarded, while samples away from the intersection point are unchanged. The algebraic properties of our estimator (i.e. its PSD guarantee) are unaffected, though the extent of the discarded region may affect bias.

    \medskip
    \item \textbf{What if the largest tangent ball is unbounded?} We follow the heuristic of \citet{yu-2024-tangent-ball} by clamping the maximum diameter for tangent balls to the shortest side length of the mesh's axis-aligned bounding box. For the exterior estimator, we follow the same heuristic, though the bounding box is taken w.r.t. the Kelvin-inverted surface.

    \medskip
    \item \textbf{How is nested geometry treated?} Nested geometry introduces a modeling ambiguity that is not resolved by the surface geometry alone. For example, if a closed component lies entirely inside another closed component, one may interpret the inner component as an internal obstacle, a cavity boundary, a separate solid object, or an artifact to be ignored, depending on the intended application. Our implementation adopts the convention that the entire input surface is absorbing: any triangle may serve as a terminating boundary for WoS samples. Thus, nested components are treated as additional boundary components of the sampled domain, rather than being discarded. This convention is appropriate when the entire surface is intended to participate in the boundary process, but it is ultimately a modeling choice; applications with different semantics may wish to remove nested components, or assign them different boundary behavior.
\end{enumerate}

\section{Point-based $k$-NN Interpolator}
\label{sec:supp_knn}
Our point-based estimators employ $k$-NN interpolators when evaluating Galerkin basis functions on the boundary---i.e., to compute $b_s=\phi(s)$ and $b_t=\phi(t)$ for a given boundary function $\phi$ (analogous to the barycentric interpolation used in Algorithm \ref{alg:interior-psd-estimator}). To reduce spurious interactions between boundary points, we use a bilateral weighting on the $k$ nearest neighbors around the query point. 

In particular, given a query point $q$, we find its $k$ nearest point cloud samples $\mathcal{N}_k(q)$. The value of a boundary function $\phi$ at $q$ is approximated by a weighted average over these neighbors,
\begin{equation}
    \phi(q)
    \approx
    \sum_{i \in \mathcal{N}_k(q)} w_i(q)\,\phi(x_i),
\end{equation}
where the bilateral weights, $w_i$, depend on both spatial distance and normal alignment,
\begin{align}
    \begin{aligned}
    \widetilde{w}_i(q)
    &=
    \exp\!\left(
        -\frac{\|q-x_i\|^2}{2\sigma^2}
    \right)
    \exp\!\left(
        -\frac{1-\langle n_q,n_i\rangle}{\tau_n}
    \right),\\
    w_i(q)
    &=
    \frac{\widetilde{w}_i(q)}
    {\sum_{j \in \mathcal{N}_k(q)} \widetilde{w}_j(q)}.
    \end{aligned}
\end{align}
Here, $x_i$ and $n_i$ are the position and normal of the $i$-th point, and $n_q$ is the normal at the query point. The parameter $\sigma$ determines spatial falloff, and the parameter $\tau_n$ determines normal falloff. In our implementation, we use $\sigma=0.025$, $\tau_n = 0.8$, and $k=8$ neighbors.

\begin{figure}
    \centering
    \includegraphics[width=\linewidth , trim=0 1.5cm 0 0]{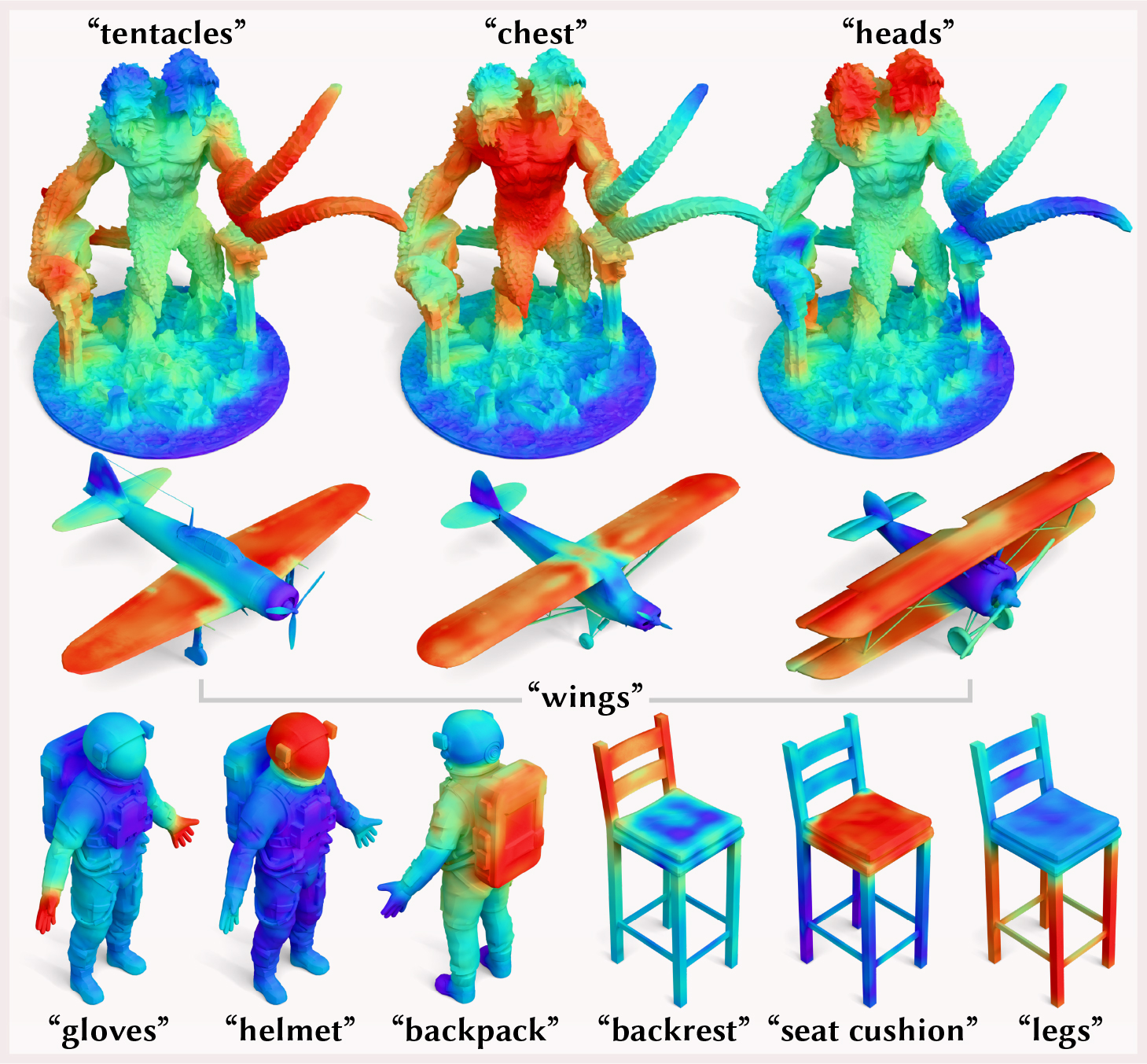}
    \caption{Unfiltered saliency maps from Steklov-CLIP, corresponding to the \emph{filtered} saliency maps in Figure \ref{fig:clip_saliency}.}
    \label{fig:clip_saliency_unfiltered}
\end{figure}

\section{Derivations for Interior and Exterior DtN Estimators}
\label{sec:supplemental}

\begin{prettyalgorithm}{alg:steklov-net-block}{Steklov Network Block}
\begin{algorithmic}
  \Require Boundary features $\*f\in\R^{N\times C}$; mass matrix $\*M$;
  Steklov eigenbases of size $k$, 
  $(\stekBasisInt,\*\lambda_{\mathrm{int}})$,
  $(\stekBasisExt,\*\lambda_{\mathrm{ext}})$
  \Ensure Updated boundary features $\*f_{\mathrm{out}}\in\R^{N\times C}$

  \State $\*f_{\mathrm{in}} \gets \*f$

  \\
  \LineComment{Apply interior/exterior spectral heat filters with learned times $\*t$}
  \State $\*h \gets$  
  \Call{SteklovHeat}{$\*f$, $\stekBasisInt$, $\*\lambda_{\mathrm{int}}$, $\stekBasisExt$, $\*\lambda_{\mathrm{ext}}$, $\*M$, $\*t$}
  \State $\*f \gets \*f + \gamma_{\mathrm{heat}}\odot \*h$
  \\

  \LineComment{Steklov-Galerkin Attention:}
  \State $\*u \gets \*f\,\*W_\mathrm{in}$

  \For{$s\in\{\mathrm{int},\mathrm{ext}\}$}
    
    \LineComment{Projection into Steklov spectrum}
    \State $\*Z_s \gets \*\Psi_s^\T \*M \*u$

    \LineComment{Project to spectral queries, keys, and values}
    \State $\*Q_s \gets \*Z_s\*W^Q_s,\quad
           \*K_s \gets \*Z_s\*W^K_s,\quad
           \*V_s \gets \*Z_s\*W^V_s$

    \LineComment{Modulate using eigenvalue-dependent spectral filters}
    \State $\*Q_s \gets$
    \Call{Modulate}{$\*Q_s,\boldsymbol\lambda_s,\boldsymbol\eta^Q_s$}
    \State $\*K_s \gets$
    \Call{Modulate}{$\*K_s,\boldsymbol\lambda_s,\boldsymbol\eta^K_s$}
    \State $\*V_s \gets$
    \Call{Modulate}{$\*V_s,\boldsymbol\lambda_s,\boldsymbol\eta^V_s$}

    \LineComment{Assemble bilinear operator from Steklov functions}
    \State $\*G_s \gets \tfrac{1}{k}\*K_s^\T \*V_s$
  \EndFor

  \LineComment{Learned interior/exterior mixing (per-head)}
  \State $\widetilde{\*G}_{\mathrm{int}}
    \gets
    \alpha_{\mathrm{ii}}\*G_{\mathrm{int}}
    +
    \alpha_{\mathrm{ie}}\*G_{\mathrm{ext}}$

  \State $\widetilde{\*G}_{\mathrm{ext}}
    \gets
    \alpha_{\mathrm{ei}}\*G_{\mathrm{int}}
    +
    \alpha_{\mathrm{ee}}\*G_{\mathrm{ext}}$

  \LineComment{Apply operators to queries}
  \State $\*Y_{\mathrm{int}}
    \gets
    \tfrac{1}{\sqrt{d}}\,
    \*Q_{\mathrm{int}}\widetilde{\*G}_{\mathrm{int}},
    \quad
    \*Y_{\mathrm{ext}}
    \gets
    \tfrac{1}{\sqrt{d}}\,
    \*Q_{\mathrm{ext}}\widetilde{\*G}_{\mathrm{ext}}$

  \LineComment{Project back to spatial domain and apply linear layer}
  \State $\*a \gets
    \left(
      \stekBasisInt\*Y_{\mathrm{int}}
      +
      \stekBasisExt\*Y_{\mathrm{ext}}
    \right)\*W_{\mathrm{out}}$

  \State $\*f \gets \*f + \gamma_{\mathrm{attn}}\odot \*a$

  \\

  \LineComment{Mix updated/original features}
  \State $\*m \gets$
  \Call{MLP}{$\left[ \*f_{\mathrm{in}}, \*f \right]$}
  \State $\*f_{\mathrm{out}} \gets \*f + \gamma_{\mathrm{mlp}}\odot \*m$

  \State \Return $\*f_{\mathrm{out}}$
    
  \\
  \Function{SteklovHeat}{$\*f,\stekBasisInt,\*\lambda_{\mathrm{int}},
    \stekBasisExt,\*\lambda_{\mathrm{ext}},\*M,\*t$}
    \LineComment{Evolve features using interior/exterior Steklov heat per-channel}

    \State $\*f_{\mathrm{int}},\*f_{\mathrm{ext}}
      \gets \Call{SplitChannels}{\*f}$
    \Comment{Each has $C/2$ channels}

    \State $\*t_{\mathrm{int}},\*t_{\mathrm{ext}}
      \gets \Call{SplitChannels}{\*t}$
    \Comment{Learned heat times per channel}

    \State $\*h^c_{\mathrm{int}} \gets \stekBasisInt \begin{bmatrix}e^{-\lambda^0_{\mathrm{int}} t_{\mathrm{int}}^c}\\... \\e^{-\lambda^k_{\mathrm{int}} t_{\mathrm{int}}^c}\end{bmatrix}\odot (\stekBasisInt^\T \*M\,\*f_{\mathrm{int}}^c)$

    \State $\*h^c_{\mathrm{ext}} \gets \stekBasisExt \begin{bmatrix}e^{-\lambda^0_{\mathrm{ext}} t_{\mathrm{ext}}^c}\\... \\e^{-\lambda^k_{\mathrm{ext}} t_{\mathrm{ext}}^c}\end{bmatrix}\odot (\stekBasisExt^\T \*M\,\*f_{\mathrm{ext}}^c)$

    \State \Return $\Call{Concat}{\*h_{\mathrm{int}},\*h_{\mathrm{ext}}}$
  \EndFunction

  \\
  \Function{Modulate}{$\*X_s,\*\lambda_s,\boldsymbol\eta^\bullet_s$}
    \LineComment{Modulate spectral function with multi-scale heat kernel}
    \State $\boldsymbol\tau \gets \{\tau_1,\ldots,\tau_L\}$
    \Comment{Fixed log-spaced heat times}

    \State $\boldsymbol\omega^\bullet_s
      \gets
      \operatorname{softmax}_{\ell}(\boldsymbol\eta^\bullet_s)$
    \Comment{Learned convex weights over heat scales}

    \State $\left[\*D^\bullet_s\right]_{kk}
      \gets
      \sum_{\ell=1}^{L}
      \omega^\bullet_{s,\ell}\exp(-\tau_\ell\lambda^s_k)$

    \State \Return $\*D^\bullet_s\*X_s$
    \EndFunction
\end{algorithmic}
\end{prettyalgorithm}

This supplemental section fills in the derivations for the interior and exterior Beurling-Deny bilinear forms used in \refsec{sec:monte-carlo}, together with the closed-form identities that underlie our samplers. Throughout this section, we translate from the conventions of \citet{chen-fukushima-2011-dtn}, who formulate the DtN operator through the Dirichlet form $(\frac{1}{2}\*D, H^1(\vol))$ with generator $\frac{1}{2}\Delta$ (here, $\*D[f,g]:=\int_\vol \nabla f\cdot\nabla g\ dV$ is the Dirichlet integral). Our convention (\refeq{eq:dirichlet-form-dtn-volume}) uses the unscaled Dirichlet form $(\*D, H^1(\vol))$ generated by $\Delta$, so equations from this source pick up a factor of $2$ when brought into our notation. Recall that our outward normals $n_s$ at boundary points $s\in\boundary$ always point \emph{out} of $\vol$ and \emph{into} $\vole$.

\subsection{Beurling-Deny Formula}
\label{sec:beurling-deny-supp}

\paragraph*{Beurling-Deny Decomposition of DtN}

\citet[Eq. 5.8.4]{chen-fukushima-2011-dtn} write the Dirichlet energy of a function $f\in H^{\frac{1}{2}}(\boundary)$ as
\begin{align}
    \label{eq:beurling-deny-supp}
    \E[f, f] &= \frac{1}{2}\iint\limits_{\boundary\times\boundary} \big(f(s)-f(t)\big)^2\J^\Omega(s,t)\ ds\ dt.
\end{align}
Applying the \emph{Polarization Identity} to \refeq{eq:beurling-deny-supp} gives the bilinear form
\begin{align}
    \label{eq:beurling-deny-bilinear-supp}
    \E[f, g] &= \frac{1}{2}\iint\limits_{\boundary\times\boundary} \big(f(s)-f(t)\big)\big(g(s)-g(t)\big)\,\J^\Omega(s,t)\ ds\ dt,
\end{align}
which is the starting point for deriving our interior estimator. A similar application of the \emph{Polarization Identity} to \cite[Eq. 5.8.9]{chen-fukushima-2011-dtn} provides the starting point for deriving our exterior estimator.

\subsection{Jump Kernel}
\label{sec:jump-kernel-supp}

\paragraph*{Jump Kernel of a Ball.} After establishing the jump kernel as the inward normal derivative of the Poisson kernel \cite[Eq. 5.8.2]{chen-fukushima-2011-dtn}, we can derive a closed-form expression for the jump measure between a point $s\in\boundary$ and another point $z$ on the boundary of a ball $B_r(c)\subset\vol$ that is tangent to the boundary at $s$. The ball has radius $r>0$, center $s-rn_s$, and Poisson kernel
\begin{align}
    P^{B_r(c)}(x\to z) = \frac{1}{4\pi r}\frac{r^2-|x-c|^2}{|x-z|^3},
\end{align}
which vanishes when $x$ lies on $\partial B_r(c)$. Therefore, we have
\begin{align}
\begin{aligned}
    \label{eq:ball-jump-kernel-supp}
    \J^{B_r(c)}(s,z) &= -\partial_{n_s}\P^{B_r(c)}(s\to z)\\
    &= \lim_{\epsilon\to 0} \frac{\P^{B_r(c)}(s-\epsilon n_s\to z)-\P^{B_r(c)}(s\to z)}{\epsilon}\\
    &= \lim_{\epsilon\to 0} \frac{\P^{B_r(c)}(s-\epsilon n_s\to z)}{\epsilon}\\
    &= \lim_{\epsilon\to 0} \frac{1}{4\pi r\epsilon}\frac{r^2-|(s-\epsilon n_s)-c|^2}{|(s-\epsilon n_s)-z|^3}\\
    &= \lim_{\epsilon\to 0} \frac{1}{4\pi r\epsilon}\frac{r^2-(r-\epsilon)^2}{|(s-\epsilon n_s)-z|^3}\\
    &= \lim_{\epsilon\to 0} \frac{1}{4\pi r\epsilon}\frac{r^2-(r^2-2r\epsilon +\epsilon^2)}{|(s-\epsilon n_s)-z|^3}\\
    &= \lim_{\epsilon\to 0} \frac{1}{4\pi r}\frac{2r -\epsilon}{|(s-\epsilon n_s)-z|^3}\\
    &= \frac{1}{4\pi r}\frac{2r}{|s-z|^3}\\
    &= \frac{1}{2\pi|s-z|^3}.
\end{aligned}
\end{align}
This closed-form ball jump kernel is the analytical building block that lets us factor WoS-style walks out of the general jump kernel.

\paragraph*{Jump Kernel of $\Omega$}
The ball jump kernel $J^{B_r(c)}$ is known in closed-form, but the jump kernel $J^\vol$ for a general domain is not. To maximize sampling efficiency, we would like to leverage the analytically known jump measure between two ball surface points in order to factor out the portion of a jump on $\boundary$ that can be handled exactly. By writing the jump kernel in terms of the Poisson kernel (\refeq{eq:jump-kernel-poisson}) and using the \emph{Strong Markov Property} (\refeq{eq:strong-markov-property}), we obtain
\begin{align}
    \label{eq:volume-jump-kernel-supp}
    \begin{aligned}
    \J^\vol(s,t) &= -\partial_{n_s} \P^\vol(s\to t) \\
    &= -\partial_{n_s} \int_{\partial B_r(c)} \P^{B_r(c)}(s\to z)\,\P^\vol(z\to t)\ dz\\
    &= \int_{\partial B_r(c)} -\partial_{n_s}\P^{B_r(c)}(s\to z)\,\P^\vol(z\to t)\ dz\\
    &= \int_{\partial B_r(c)} \J^{B_r(c)}(s, z)\,\P^\vol(z\to t)\ dz.
    \end{aligned}
\end{align}
The resulting factorization reveals that the singular, differential portion of $J^\vol$ is fully captured by the known ball jump kernel.

\subsection{Kelvin Transform}
\label{sec:kelvin-transform-supp}

The \emph{Kelvin transform} is a conformal mapping $\kelvin: \R^3\cup\{\infty\}\to\R^3\cup\{\infty\}$ sending $x\mapsto x/|x|^2$ (with $0\mapsto\infty$ and $\infty\mapsto 0$). Assuming $0\in\vol$ (which can always be arranged by translation), we define 
\begin{align}
    \kelvin{\vole}:=\kelvin(\vole\cup\{\infty\})
\end{align}
as the inverted exterior. Under the Kelvin transform, $\kelvin{\vole}$ is a \emph{bounded} domain in $\R^3$ that includes the origin (as the image of $\infty$). We denote the image of $x$ under the Kelvin transform as $\kelvin{x}:=\kelvin(x)$. It is important to recognize that $\kelvin$ is its own inverse, so that $\kelvin(\kelvin{x})=x$.

\paragraph{Jacobian of the Kelvin Transform} The $ij$-th entry of the Kelvin transform's Jacobian is
\begin{align*}
    \*J_{\kelvin}(x)_{ij} &= \frac{\partial \kelvin{x}_i}{\partial x_j} = \frac{\partial \left(x_i/|x|^2\right)}{\partial x_j}.
\end{align*}
Applying the \emph{Quotient Rule}, we have that $\*J_{\kelvin}(x)_{ij}$ is equal to
\begin{align*}
     \frac{|x|^2 \frac{\partial x_i}{\partial x_j}-x_i\frac{\partial \sum x_i^2}{\partial x_j}}{|x|^4}= \frac{\delta_{ij}|x|^2-2x_ix_j}{|x|^4}= \frac{1}{|x|^2}\left(\delta_{ij}-\frac{2x_ix_j}{|x|^2}\right).
\end{align*}
In matrix form, the Jacobian simplifies to
\begin{align}
    \label{eq:kelvin-transform-jacobian}
    \*J_{\kelvin}(x) = \frac{1}{|x|^2}\left(\*I-\frac{2xx^\T}{|x|^2}\right) = \frac{1}{|x|^2}\*H_{\hat{x}},
\end{align}
where $\*H_{\hat{x}}$ is the \emph{Householder reflection} about the direction $\hat{x}=x/|x|$. 

\paragraph{Boundary Scaling} Based on its Jacobian, the linear scale factor of $\kelvin$ is $\ell(x)=1/|x|^2$. Surface area elements on a 2D manifold scale by $\ell(x)^2$, which means
\begin{align}
    \label{eq:inverted-surface-area}
    d\sigma(\kelvin{s}) = \frac{1}{|s|^4}d\sigma(s)\quad\text{and}\quad d\sigma(s) = \frac{1}{|\kelvin{s}|^4}d\sigma(\kelvin{s}).
\end{align}
In general, this means we can transform a surface integral on the inverted boundary $\kelvin{\boundary}$ into an integral on the primal boundary $\boundary$ by introducing the appropriate density-correcting weight $|s|^{-4}$.

\paragraph{Normal Inversion} Applying the Kelvin transform's Jacobian (\refeq{eq:kelvin-transform-jacobian}) to $n_s$ reflects the normal about the radial axis $\hat{s}$ and introduces a local scaling term,
\begin{align}
    \label{eq:jacobian-times-normal-supp}
    \*J_\kappa(s)\,n_s &= \frac{1}{|s|^2}\,\*H_{\hat{s}}\,n_s = \frac{1}{|s|^2} \left(-n_{\kelvin{s}}\right) = -|\kelvin{s}|^2 n_{\kelvin{s}},
\end{align}
where $n_{\kelvin{s}}$ denotes the unit normal---pointing out of $\kelvin{\vole}$ and into $\kelvin{\vol}$---defined at points $\kelvin{s}\in\kelvin{\boundary}$ on the inverted boundary.

\paragraph{Kelvin Transform of a Function} Let $u$ be a function defined on $\vole\subset \R^3\setminus\{0\}$. Then, the function $\kelvin[u]$ defined on $\kelvin{\vole}\setminus\{0\}$ by
\begin{align}
    \label{eq:kelvin-transform-of-function}
    \kelvin[u](\kelvin{x}) := |x|\,u(x)
\end{align}
is the \emph{Kelvin transform of $u$}. In particular, if $u$ is harmonic on $\vole$ and satisfies $u(x)\to 0$ as $x\to\infty$, then the transformation $\kappa[u]$ is harmonic on $\kelvin{\vole}$ \cite{axler-2006-harmonic}.

\subsection{Exterior Poisson Kernel}
\label{sec:exterior-poisson-kernel-supp}

\paragraph{Kelvin Transform of the Poisson Kernel} We wish to relate the Poisson kernels of the two harmonic functions given in \refeq{eq:kelvin-transform-of-function} (namely $u$ and $\kelvin[u]$). Suppose $u=\H^{\vole}[g]$ for some boundary data $g:\boundary\to\R$. By \refeq{eq:poisson-kernel-analytic}, we obtain the Poisson integral
\begin{align}
    u(x) = \int_{\boundary} g(s)\,\P^{\vole}(x\to s)\ ds
\end{align}
for $x\in\vole$. Then, $\kelvin[u]$ solves the \emph{bounded} Dirichlet problem on $\kelvin{\vole}$ with boundary conditions $\kelvin[u](\kelvin{s})=|s|\,g(s)$ for $\kelvin{s}\in\kelvin{\boundary}$. Its Poisson integral takes the form
\begin{align}
    \kelvin[u](\kelvin{x}) = \int_{\kelvin{\boundary}} |s|\,g(s)\,\P^{\kelvin{\vole}}(\kelvin{x}\to\kelvin{s})\ d\kelvin{s},
\end{align}
this time for $\kelvin{x}\in\kelvin{\vole}$. We can represent this as a boundary integral on the primal surface $\boundary$ by introducing a $|s|^{-4}$ scale factor (\refeq{eq:inverted-surface-area}),
\begin{align}
    \kelvin[u](\kelvin{x}) = \int_{\boundary} |s|^{-3}\,g(s)\,\P^{\kelvin{\vole}}(\kelvin{x}\to\kelvin{s})\ ds.
\end{align}
By definition (\refeq{eq:kelvin-transform-of-function}), $\kelvin[u](\kelvin{x}) = |x|\,u(x)$, which implies that
\begin{align*}
    \int_{\boundary} |s|^{-3}\,g(s)\,\P^{\kelvin{\vole}}(\kelvin{x}\to\kelvin{s})\ ds &= |x|\int_{\boundary} g(s)\,\P^{\vole}(x\to s)\ ds.
\end{align*}
Since $|x|^{-1}=|\kelvin{x}|$, the $|x|$ factor can be moved to the left-hand side,
\begin{align*}
    \int_{\boundary} g(s)\,|\kelvin{x}|\,|\kelvin{s}|^{3}\,\P^{\kelvin{\vole}}(\kelvin{x}\to\kelvin{s})\ ds &= \int_{\boundary} g(s)\,\P^{\vole}(x\to s)\ ds.
\end{align*}
Here, $g$ is an arbitrary function. Because the two integrals must agree pointwise for any choice of $g$, we reach the identity 
\begin{align}
    \label{eq:poisson-kernel-kelvin}
    \P^{\vole}(x\to s) &= |\kelvin{x}|\,|\kelvin{s}|^3\,\P^{\kelvin{\vole}}(\kelvin{x}\to\kelvin{s})
\end{align}
relating the Poisson kernel of the exterior domain $\vole$ with that of the inverted domain $\kelvin{\vole}$. This identity underpins our exterior DtN estimator formulation, which exploits the boundedness of $\kelvin{\vole}$.

\subsection{Exterior Jump Kernel}
\label{sec:exterior-jump-kernel-supp}

Like the interior case (\refeq{eq:jump-kernel-poisson}), the exterior jump kernel is the normal derivative of the exterior Poisson kernel,
\begin{align}
    \label{eq:exterior-jump-kernel-poisson-supp}
    \J^{\vole}(s,t) &= \partial_{n_s}\,\P^{\vole}(s\to t).
\end{align}
Substituting the exterior Poisson kernel relation in \refeq{eq:poisson-kernel-kelvin},
\begin{align*}
\begin{aligned}
    \J^{\vole}(s,t) &= \partial_{n_s}\left[|\kelvin{s}|\,|\kelvin{t}|^3\,\P^{\kelvin{\vole}}(\kelvin{s}\to\kelvin{t})\right]\\
    &= \nabla_s\left[|\kelvin{s}|\,|\kelvin{t}|^3\,\P^{\kelvin{\vole}}(\kelvin{s}\to\kelvin{t})\right]\cdot n_s\\
    &= |\kelvin{t}|^3\left[\,\P^{\kelvin{\vole}}(\kelvin{s}\to\kelvin{t})\,\nabla_s\,|\kelvin{s}| + |\kelvin{s}|\,\nabla_s\,\P^{\kelvin{\vole}}(\kelvin{s}\to\kelvin{t})\right]\cdot n_s.
\end{aligned}
\end{align*}
The first term vanishes because $\P^{\kelvin{\vole}}(\kelvin{s}\to\kelvin{t}) = 0$, leaving
\begin{align}
    \label{eq:exterior-jump-prefactor-split}
    \J^{\vole}(s,t) = |\kelvin{s}|\,|\kelvin{t}|^3\,\nabla_s\,\P^{\kelvin{\vole}}(\kelvin{s}\to\kelvin{t})\cdot n_s.
\end{align}
By the chain rule,
\begin{align}
\begin{aligned}
    \nabla_s\,\P^{\kelvin{\vole}}(\kelvin{s}\to\kelvin{t})\cdot n_s &= \*J_{\kelvin}(s)^\T\,\nabla_{\kelvin{s}}\,\P^{\kelvin{\vole}}(\kelvin{s}\to\kelvin{t})\cdot n_s\\
    &= \nabla_{\kelvin{s}}\,\P^{\kelvin{\vole}}(\kelvin{s}\to\kelvin{t})\cdot\*J_{\kelvin}(s)\,n_s\\
    &= \nabla_{\kelvin{s}}\,\P^{\kelvin{\vole}}(\kelvin{s}\to\kelvin{t})\cdot\left(-|\kelvin{s}|^2\,n_{\kelvin{s}}\right)\\
    &= -|\kelvin{s}|^2\,\nabla_{\kelvin{s}}\,\P^{\kelvin{\vole}}(\kelvin{s}\to\kelvin{t})\cdot n_{\kelvin{s}}\\
    &= -|\kelvin{s}|^2\,\partial_{n_{\kelvin{s}}}\,\P^{\kelvin{\vole}}(\kelvin{s}\to\kelvin{t})\\
    &= |\kelvin{s}|^2\,\J^{\kelvin{\vole}}(\kelvin{s},\kelvin{t}),
\end{aligned}
\end{align}
where the third equality applies the image of the surface normal under Kelvin inversion (\refeq{eq:jacobian-times-normal-supp}) and the final equality applies the interior jump kernel definition (\refeq{eq:jump-kernel-poisson}) on the bounded domain $\kelvin{\vole}$. Assembling, we attain
\begin{align}
    \label{eq:exterior-jump-kernel-final-supp}
    \J^{\vole}(s,t) &= |\kelvin{s}|^3\,|\kelvin{t}|^3\,\J^{\kelvin{\vole}}(\kelvin{s},\kelvin{t}).
\end{align}
The exterior jump kernel on the unbounded domain $\vole$ is---up to a pair of distortion-correcting scale factors---the interior style jump kernel on the bounded inverted domain $\kelvin{\vole}$. Notably, $\J^{\kelvin{\vole}}$ admits the very same tangent-ball decomposition which we used to reduce variance in the interior case (\refeq{eq:volume-jump-kernel}).

\subsection{Killing Measure}
\label{sec:killing-measure-supp}

\paragraph{Escape Probability}
The escape probability $q:\vole\to[0,1]$ on exterior points is defined as
\begin{align}
    \label{eq:escape-probability-supp}
    q(x) &:= 1-\H^{\vole}[\*1](x),
\end{align}
which is harmonic on the unbounded exterior domain $\vole$, vanishes on $\boundary$, and approaches $1$ as $|x|\to\infty$ \cite[Eq. 5.8.6]{chen-fukushima-2011-dtn}. We claim that $q(x)$ is equivalent to
\begin{align}
    \label{eq:escape-probability-candidate}
    \tilde{q}(x):=-4\pi\,|\kelvin{x}|\,G^{\kelvin{\vole}}(\kelvin{x},0),
\end{align}
which is defined on the unbounded exterior domain $\vole\cup\{\infty\}$. Here, $G^{\kelvin{\vole}}$ is the \emph{Green's function} of the inverted domain, given by
\begin{align}
    \label{eq:greens-split}
    G^{\kelvin{\vole}}(\kelvin{x}, 0) := -\frac{1}{4\pi|\kelvin{x}|} + u(\kelvin{x}, 0),
\end{align}
for some $u$ that is harmonic on $\kelvin{\vole}$ \cite[Exercise 8.3.2]{stakgold-2011-greens-func}. After substituting \refeq{eq:greens-split} into \refeq{eq:escape-probability-candidate}, we obtain
\begin{align}
\label{eq:escape-probability-difference}
\begin{aligned}
    \tilde{q}(x) &= 1-4\pi\,|\kelvin{x}|\,u(\kelvin{x},0).
\end{aligned}
\end{align}
There are two things to notice here:

\begin{enumerate}
    \medskip
    \item \textbf{Boundary conditions.} $G^{\kelvin{\vole}}(\kelvin{x},0)$ vanishes on $\kelvin{\boundary}$, which means that $\tilde{q}(x)$ also vanishes on $\boundary$---matching the behavior of $q(x)$ on the boundary.

    \medskip
    \item \textbf{Limit at infinity.} Because $u(\kelvin{x}, 0)$ is harmonic on $\kelvin{\vole}$, which includes $0$, it is bounded on a neighborhood of $\kelvin{x}=0$. Hence,
    \begin{align}
        \label{eq:greens-func-zero-limit}
        \lim_{\kelvin{x}\to 0} 4\pi\,|\kelvin{x}|\,u(\kelvin{x},0) = 0.
    \end{align}
    As $|x|\to\infty$, we have $|\kelvin{x}|\to 0$. By   \refeq{eq:escape-probability-difference} and \refeq{eq:greens-func-zero-limit}, we get
    \begin{align}
        \lim_{x\to\infty} \tilde{q}(x) = 1 - 0 = 1,
    \end{align}
    which agrees with $q(x)$ at infinity.
\end{enumerate}

\medskip\noindent
By the uniqueness of exterior Dirichlet problem solutions, we must conclude that $q(x)=\tilde{q}(x)$. In other words, escape probability in the exterior domain $\vole$ is equivalently written in terms of the Green's function of the Laplacian in the inverted domain $\kelvin{\vole}$, such that
\begin{align}
    \label{eq:escape-probability-final}
    q(x) = -4\pi\,|\kelvin{x}|\,G^{\kelvin{\vole}}(\kelvin{x},0).
\end{align}
This re-characterization will become vital for deriving an expression for the killing measure $\K^{\vole}$ that can actually be sampled.

\paragraph{Sampling Killing Measure} The starting point for our killing measure estimator comes from \citet[Eq. 5.8.6]{chen-fukushima-2011-dtn},
\begin{align}
    \label{eq:killing-measure-normal-derivative-supp}
    \K^{\vole}(s) &= \partial_{n_s} q(s).
\end{align}
Substituting in our work from the preceeding section (\refeq{eq:escape-probability-final}),
\begin{align*}
    \begin{aligned}
    \K^{\vole}(s) &= \partial_{n_s}\left(-4\pi\,|\kelvin{s}|\,G^{\kelvin{\vole}}(\kelvin{s},0)\right)\\
    &= \nabla_s\left(-4\pi\,|\kelvin{s}|\,G^{\kelvin{\vole}}(\kelvin{s},0)\right)\cdot n_s\\
    &= -4\pi\left[G^{\kelvin{\vole}}(\kelvin{s},0)\,\nabla_s\,|\kelvin{s}|+|\kelvin{s}|\,\nabla_s \,G^{\kelvin{\vole}}(\kelvin{s},0)\right]\cdot n_s.
    \end{aligned}
\end{align*}
The first term vanishes because $G^{\kelvin{\vole}}(\kelvin{s},0)=0$ when $\kelvin{s}\in\kelvin{\boundary}$ ($G^{\kelvin{\vole}}$ satisfies homogeneous Dirichlet boundary conditions). Then,
\begin{align}
    \label{eq:killing-measure-product-rule-simplified}
    \K^{\vole}(s) &= -4\pi\,|\kelvin{s}|\,\nabla_s \,G^{\kelvin{\vole}}(\kelvin{s},0) \cdot n_s.
\end{align}
Using the chain rule,
\begin{align}
\begin{aligned}
    \nabla_s \,G^{\kelvin{\vole}}(\kelvin{s},0) \cdot n_s &= \,\*J_{\kelvin}(s)^\T\,\nabla_{\kelvin{s}}\,G^{\kelvin{\vole}}(\kelvin{s},0) \cdot n_{s} \\
    &= \nabla_{\kelvin{s}}\,G^{\kelvin{\vole}}(\kelvin{s},0) \cdot\*J_\kappa(s)\,n_s \\
    &= \nabla_{\kelvin{s}} \,G^{\kelvin{\vole}}(\kelvin{s},0) \cdot \left(-|\kelvin{s}|^2\,n_{\kelvin{s}}\right)\\
    &= -|\kelvin{s}|^2\,\nabla_{\kelvin{s}} \,G^{\kelvin{\vole}}(\kelvin{s},0) \cdot  n_{\kelvin{s}}\\
    &= -|\kelvin{s}|^2\,\partial_{n_{\kelvin{s}}}G^{\kelvin{\vole}}(\kelvin{s},0)\\
    &= -|\kelvin{s}|^2\,\partial_{n_{\kelvin{s}}}G^{\kelvin{\vole}}(0,\,\kelvin{s})\\
    &= -|\kelvin{s}|^2\,\P^{\kelvin{\vole}}(0\to\kelvin{s}),
\end{aligned}
\end{align}
where the third equality follows from the Kelvin transform of surface normals (\refeq{eq:jacobian-times-normal-supp}), the sixth equality holds because the Dirichlet Green's function is symmetric in its two arguments, and the last equality identifies the normal derivative of the Green's function as the Poisson kernel (\refeq{eq:poisson-kernel-to-greens-func}). Substituting this back into \refeq{eq:killing-measure-product-rule-simplified},
\begin{align}
    \label{eq:killing-measure-final-supp}
    \K^{\vole}(s) &= 4\pi\,|\kelvin{s}|^3\,\P^{\kelvin{\vole}}(0\to\kelvin{s}),
\end{align}
which remarkably implies that we can sample the killing measure by starting random walks from the inversion center (i.e. $\kelvin(\infty)$, which is defined to be the origin).

\subsection{Exterior DtN Estimator}
\label{sec:exterior-dtn-estimator-supp}

Let $\{\phi_k\}_{k=1}^{K}$ be our chosen boundary basis, let $b(s)\in\R^K$ be a pointwise evaluation of the basis at $s\in\boundary$, and let $d(s,t):=b(s)-b(t)\in\R^K$ denote the difference of basis evaluations at two different boundary points. Using the Beurling Deny form given in \refeq{eq:exterior-dtn-beurling-deny-form}, the reduced \emph{exterior} Steklov matrix $\stiffnesse\in\R^{K\times K}$, with entries equal to $(\stiffnesse)_{ij}=\E[\phi_i,\phi_j]$, is expressed as
\begin{align}
    \label{eq:exterior-steklov-matrix-beurling-deny-supp}
    \begin{aligned}
    \stiffnesse &= \frac{1}{2}\iint\limits_{\boundary\times\boundary} d(s,t)\,d(s,t)^\T\J^{\vole}(s,t)\ ds\ dt\\
    &\quad+\int_{\boundary} b(s)\,b(s)^\T\,\K^{\vole}(s)\ ds.
    \end{aligned}
\end{align}
Our goal is to trace the formulation of the \emph{interior} PSD-by-construction estimator of the DtN operator (\refsec{sec:improved-beurling-deny-estimator}), but using the Kelvin-transformed proxies for the exterior jump kernel (\refeq{eq:exterior-jump-kernel-final-supp}) and killing measure (\refeq{eq:killing-measure-final-supp}) instead. Mirroring the tangent-ball decomposition in \refsec{sec:jump-kernel-supp}, the exterior domain jump kernel factors as
\begin{align*}
    \J^{\vole}(s,t)&=|\kelvin{s}|^3\,|\kelvin{t}|^3\int_{\partial B_r(\kelvin{c})} \J^{B_r(\kelvin{c})}(\kelvin{s}, \kelvin{z})\,\P^{\kelvin{\vole}}(\kelvin{z}\to \kelvin{t})\ d\kelvin{z}.
\end{align*}
After substituting this quantity into the jump term of \refeq{eq:exterior-steklov-matrix-beurling-deny-supp}, and substituting \refeq{eq:killing-measure-final-supp} into its killing term, we can follow the same rearrangement that precedes \refeq{eq:stiffness-matrix-final-integral}, yielding
\begin{align}
    \label{eq:exterior-stiffness-rearranged-supp}
    \begin{aligned}
    \stiffnesse &= \frac{1}{2}\mkern-16mu\iint\limits_{\mskip24mu\boundary\times\partial B_r(\kelvin{c})}\mkern-16mu|\kelvin{s}|^3\,\J^{B_r(\kelvin{c})}(\kelvin{s},\kelvin{z})\\[-2pt]
    &\mskip36mu\cdot\left(\int_{\boundary}d(s,t)\,d(s,t)^\T\,|\kelvin{t}|^3\,\P^{\kelvin{\vole}}(\kelvin{z}\to\kelvin{t})\ dt\right)\ d\kelvin{z}\ ds\\
    &+\bigg(\int_{\boundary} b(t)\,b(t)^\T\,4\pi\,|\kelvin{t}|^3\,\P^{\kelvin{\vole}}(0\to\kelvin{t})\ dt\bigg).
    \end{aligned}
    \raisetag{18pt}
\end{align}
The variable of integration of the two parenthesized boundary integrals is $t\in\boundary$ on the primal surface. However, both integrals contain the Poisson kernel of the Kelvin-transformed domain $\kelvin{\vole}$. For consistency, we can switch the variable of integration to $\kelvin{t}\in\kelvin{\boundary}$ on the inverted boundary using the relation given in \refeq{eq:inverted-surface-area},
\begin{align}
    \label{eq:exterior-stiffness-rearranged-transformed-supp}
    \begin{aligned}
    \stiffnesse &= \frac{1}{2}\mkern-16mu\iint\limits_{\mskip24mu\boundary\times\partial B_r(\kelvin{c})}\mkern-16mu|\kelvin{s}|^3\,\J^{B_r(\kelvin{c})}(\kelvin{s},\kelvin{z})\\[-2pt]
    &\mskip36mu\cdot\left(\int_{\kelvin{\boundary}}d(s,t)\,d(s,t)^\T\,\frac{1}{|\kelvin{t}|}\,\P^{\kelvin{\vole}}(\kelvin{z}\to\kelvin{t})\ d\kelvin{t}\right)\ d\kelvin{z}\ ds\\
    &+\bigg(\int_{\kelvin{\boundary}} b(t)\,b(t)^\T\,\frac{4\pi}{|\kelvin{t}|}\,\P^{\kelvin{\vole}}(0\to\kelvin{t})\ d\kelvin{t}\bigg).
    \end{aligned}
    \raisetag{18pt}
\end{align}
Now, both parentheticals are Poisson integrals (\refeq{eq:poisson-kernel-analytic}) and can be converted into expectations (\refeq{eq:poisson-kernel-probabilistic}),
\begin{align}
    \label{eq:exterior-stiffness-rearranged-expectation-supp}
    \begin{aligned}
    \stiffnesse &= \frac{1}{2}\mkern-16mu\iint\limits_{\mskip24mu\boundary\times\partial B_r(\kelvin{c})}\mkern-16mu \Exp_{\kelvin{t}}\left[d(s,t)\,d(s,t)^\T\frac{1}{|\kelvin{t}|}\right]|\kelvin{s}|^3\,\J^{B_r(\kelvin{c})}(\kelvin{s},\kelvin{z})\ d\kelvin{z}\ ds\\
    &\quad+\Exp_{\kelvin{t_0}}\left[b(t_0)\,b(t_0)^\T\,\frac{4\pi}{|\kelvin{t_0}|}\right],
    \end{aligned}
    \raisetag{18pt}
\end{align}
where the terminal WoS exit points  $\kelvin{t}\sim\omega^{\kelvin{\vole}}_{\kelvin{z}}$ and $\kelvin{t_0}\sim \omega_0^{\kelvin{\vole}}$ are sampled in the inverted domain. At last, by sampling surface points $s\sim U(\boundary)$ on the primal boundary and $\kelvin{z}\sim U(\partial B_r(\kelvin{c}))$ on the inverted tangent ball surface, we can write an \emph{exterior} estimator that operates in the Kelvin-transformed domain,
\begin{align}
\begin{aligned}
    \stiffness_{\text{ext}}
    &= \ExpUnderIcy{s,\,\kelvin{z},\,\kelvin{t}}
    \Biggl[
        \frac{|\boundary|\,|\partial B_r(\kelvin{c})|\,|\kelvin{s}|^3}{2\,|\kelvin{t}|}\,
        d(s,t)\,d(s,t)^\T\,\J^{B_r(\kelvin{c})}(\kelvin{s}, \kelvin{z})
    \Biggr]\\
    &\ \ \ \,+ \ExpUnderIcy{\kelvin{t_0}}
    \Biggl[\frac{4\pi}{|\kelvin{t_0}|}\,b(t_0)\,b(t_0)^{\T}\Biggr].
\end{aligned}
\raisetag{18pt}
\end{align}